\documentclass[11pt]{article}
\usepackage{graphicx}

\def \bx{{\bf x}}

\def \bj{{\bf j}} 
\def \bt{{\bf t}}
\def \bn{{\bf n}}
\def \bx{{\bf x}}

\def \Bb{{\bf B}}

\def \Eb{{\bf E}}
\def \Hb{{\bf H}}
\def \Sb{{\bf S}}
\def \be{\begin{equation}} 
\def \ee{\end{equation}} 
\def \d{\partial} 
\def \aa{\alpha} 
\def \bb{\beta} 
\def \dd{\delta} 
\def \gg{\gamma} 
\def \ss{\sigma} 
\def \ll{\lambda} 
\def \om{\omega}

\def \eps{\epsilon}
\def \ka{\kappa}
\def \DD{\Delta}
\def \SS{\Sigma} 
\def \LL{\Lambda}

\def \Om{\Omega}
\def \Hc{{\cal H}}

\def \Mc{{\cal M}}
\def \Ac{{\cal A}}

\def \Tc{{\cal T}}
\def \la{\langle} 
\def \ra{\rangle} 
\def \fr{\frac} 
\def \xib{{\bar \xi}}

\def \xb{{\bar x}}
\def \yb{{\bar y}}
\def \fb{{\bar f}}
\def \wg{\wedge} 

\begin{document} 
\title{Stochastic Processes, Slaves and Supersymmetry } 
\author{I.T. Drummond\thanks{email: itd@damtp.cam.ac.uk}~ 
    and R.R. Horgan\thanks{email: rrh@damtp.cam.ac.uk} \\ 
            Department of Applied Mathematics and Theoretical Physics\\ 
            Centre for Mathematical Sciences\\ 
            Wilberforce Road\\ Cambridge\\ England, CB3 0WA
        }
\maketitle
\abstract{
We extend the work of T\v{a}nase-Nicola and Kurchan on the structure of
diffusion processes and the associated supersymmetry algebra by examining
the responses of a simple statistical system to external disturbances of various kinds. 
We consider both the stochastic differential equations (SDEs) for the process and the associated 
diffusion equation.  The influence of the disturbances can be understood by augmenting 
the original SDE with an equation for {\it slave variables}. The evolution of the slave variables
describes the behaviour of line elements carried along in the stochastic flow. These line
elements together with the associated surface and volume elements constructed from them
provide the basis of the supersymmetry properties of the theory.
For ease of visualisation, and in order to emphasise a helpful electromagnetic analogy, we work in 
three dimensions. The results are all generalisable to higher dimensions and can be specialised 
to one and two dimensions. The electromagnetic analogy is a useful starting point for calculating 
asymptotic results at low temperature that can be compared with direct numerical evaluations.
We also examine the problems that arise in a direct numerical simulation of the stochastic 
equation together with the slave equations. We pay special attention to the dependence of the 
slave variable statistics on temperature. We identify in specific models the critical temperature 
below which the slave variable distribution ceases to have a variance and consider the effect on 
estimates of susceptibilities.

}

\vfill
DAMTP-2011-42

\section{\bf Introduction}

Stochastic processes have applications in many areas of physics. They are a natural way
of describing the effect of thermal interactions on dynamical systems subject to force 
fields of various kinds. These force fields could be, for example, the inter-atomic
potentials represented by an energy landscape appropriate to a molecular model \cite{WALES}. 
In another type of application, stochastic processes provide a method for achieving a
desired probability distribution for the system. This method is the basis for the 
theory of stochastic quantisation in quantum field theory. It has been exploited in the 
numerical simulation of quantum field theories \cite{PARISI,DDH,CDH}.

One way of  investigating the structure of a dynamical system is to subject
it to external disturbances. This can be accommodated in the stochastic
dynamics by inducing small changes in the motion of the particle either by changing its initial 
position or by altering the force field guiding the 
particle throughout its motion. If these external influences are infinitesimal the result, for a given
sample from the noise ensemble, is an infinitesimal shift in the path of the particle.
It can be regarded as an infinitesimal line element carried by the particle along the 
path of the original motion. Because the equation describing the evolution of this line 
element is derived from the original stochastic equation and turns out not to contain 
the noise term we refer to it as a {\it slave equation} and to the infinitesimal line element 
as a {\it slave variable}.

The response of the mean position of the system to external influences, that is its 
susceptibility, can be calculated once the statistical properties of the slave variables are 
known. Some aspects of slave variable statistics in a one dimensional case have been investigated 
in \cite{DEAN}.  In higher dimensions there exists also the possibility of examining the evolution 
and statistical properties of infinitesimal area and volume elements that can be constructed from 
the the line elements. The evolution of area and volume elements plays a significant role in 
turbulent fluid flow and in magnetohydrodynamics \cite{MOFFATT,DRUHORG,POPE,DRU1,DRU2,DRU3,SCH}. In the 
models studied in this paper these higher dimensional area and volume elements can be thought of as
the images of area and volume elements in the parameter space of the external fields. Their
statistical properties thus give rise to {\it higher dimensional susceptibilities}. These 
generalised susceptibilities can be readily computed in numerical simulations and their 
evaluation can help to elucidate the topography of the landscape function. In addition we will 
find that the hierarchy of generalised susceptibilities fits naturally into the supersymmetric 
structure underlying the stochastic dynamical system. In turn the supersymmetric analysis makes 
possible the calculation of these quantities.

A comprehensive review of material that is relevant 
particularly to one-dimensional systems and the supersymmetry approach we adopt here
is contained in reference \cite{JUNKER}. Witten \cite{WITTEN} has explained the relationship
between supersymmetric quantum mechanics and the topology of manifolds. Making use of the 
idea that diffusion processes can be regarded as quantum mechanics in imaginary time, T\v{a}nase-Nicola and 
Kurchan \cite{KURCHAN1,KURCHAN2} have developed these ideas emphasising the diffusion 
point of view and using it to construct practical simulation techniques for picking out
the saddle point structure of the landscape function. 

In this paper we extend the work of T\v{a}nase-Nicola and Kurchan \cite{KURCHAN1,KURCHAN2} 
and show it can be developed to compute significant physical observables. We do this in
part by developing an electromagnetic analogy based in the Principle of Minimum Dissipation.
This allows us to understand the structure of the physical states of the system in a way that is
particularly revealing in the low temperature limit. It can also be the basis of an optimised
approximation scheme at higher temperatures though we do not pursue this idea here.

The structure of the paper is as follows. In section \ref{basic} we describe the basic stochastic model
that we study. In section \ref{slave} we derive the slave equations, first for 
an infinitesimal change in the initial position of the particle and second for a weak external 
field. We show how they can be encoded in the supersymmetric 
formalism by introducing an appropriate supersymmetric Hamiltonian. Section \ref{simul} contains
a brief exposition of the numerical method we use for simulating the stochastic differential
and slave equations.
In section \ref{ssform} we give, for completeness and to establish notation, a brief account of the 
state space on which the Hamiltonian acts and an outline of the structure of its eigenstates.
In section \ref{emaganal}, we develop an electromagnetic analogy that helps us to understand 
the structure of the eigenstates of the supersymmetric Hamiltonian and their time evolution
under various circumstances. The analogy is developed
in section \ref{mindiss} where we show that the evolution of the eigenstates obeys a Principle
of Minimum Dissipation. 

In section \ref{low_T}, the low temperature limit in which it is possible to
compute low lying eigenvalues of the supersymmetric Hamiltonian and associated decay exponents is discussed.
The application of the supersymmetric theory to the case of an external electric 
field together with a derivation of the associated Einstein Relation is set out in section \ref{einstein}. 
Explicit examples of the calculations, in one and two dimensions, are exhibited in 
sections \ref{onedex} and \ref{twodex}, where we compare the results of various numerical techniques.  
Finally we we discuss the results and their potential
significance in section \ref{conc}.

\section{\label{basic}Basic Stochastic Model}

In this paper we consider a particle with three degrees of freedom, 
$\{X\}=\{X_1,X_2,X_3\}$, that obeys the stochastic
differential equation of the form
\be 
{\dot{X}}_n=-\phi_n(X)+W_n(t)~~,
\label{diff1}
\ee
where $W_n(t)$ is a stochastic process for which
\be
\la W_n(t)W_m(t')\ra=\dd_{nm}G(t-t')~~,
\ee
for some correlator $G(t-t')$~. The angle brackets denote averaging
over the noise ensemble. Aspects of our discussion
are independent of the precise nature of the noise $W_n(t)$~.
Ultimately however we will be interested in the case of white noise.
From now on therefore we will assume that
\be
\la W_n(t)W_m(t')\ra=2T\dd_{nm}\dd(t-t')~~.
\label{stoch}
\ee
We will refer to $T$ as the temperature of the system. 
It is implicit in our assumptions that the ensemble of
white noise processes is rotationally invariant in the
particle configuration space. While this is not the most general
case it is sufficient for illustrating our approach.

The drift term in eq(\ref{diff1}) is assumed, as indicated above, to be
derived from an energy landscape function $\phi(X)$~. That is $\phi_n(X)=\d_n\phi(X)$~.
Higher derivatives of $\phi(X)$ will also be indicated with suffixes, thus
$\phi_{mn}(X)=\d_m\d_n\phi(X)$, and so on. It is convenient and not really restrictive 
to assume that $\phi(X)$ rises like a power of $|X|$ as $|X|\rightarrow\infty$~.
As a consequence the particle will remain confined to an essentially finite region in $X$-space
throughout its motion. If we assume that at time $t=0$ the particle is at the point $\{y_n\}$
then at later time $t$, for a given element $\{\eta_n(t)\}$ of the noise ensemble,
the motion of the particle will be that solution, $\{X_n(t)\}$, of eq(\ref{diff1})
that satisfies $\{X_n(0)\}=\{y_n\}$~. The probability distribution for the position
is
\be
P(x,t)=\la\dd(x-X(t))\ra~~.
\ee
It is standard theory that $P(x,t)$ satisfies 
\be
\fr{\d}{\d t}P(x,t)=T\d_n(\d_n+\bb\phi_n(x))P(x,t)~~,
\label{diff2}
\ee
where $\bb=1/T$~. Our assumption about the form of $\phi(X)$ implies that
as $t\rightarrow\infty$ $P(x,t)\rightarrow P_0(x)$ where
\be
T\d_n(\d_n+\bb\phi_n(x))P_0(x)=0~~.
\ee
The static solution is unique and has the form
\be
P_0(x)=\fr{1}{Z}e^{-\bb\phi(x)}~~,
\label{static1}
\ee
where
\be
Z=\int dxe^{-\bb\phi(x)}~~.
\ee
Here for compactness we use $dx=dx_1dx_2dx_3$~.
Clearly
\be
(\d_n+\bb\phi_n(x))P_0(x)=0~~.
\label{ann1}
\ee
It is also convenient to introduce
\be
p_0(x)=\sqrt{P_0(x)}=\fr{1}{\sqrt{Z}}e^{-\fr{1}{2}\bb\phi(x)}~~.
\label{static2}
\ee
We have
\be
(\d_n+\fr{1}{2}\phi_n(x))p_0(x)=0~~.
\label{ann2}
\ee
Eq(\ref{diff2}) incorporates the operator ${\hat H}_0$ where
\be
{\hat H}_0=-\d_n(\d_n+\bb\phi_n(x))~~.
\label{ham1}
\ee
It is convenient to construct an associated hermitian operator $H_0$
where
\be
H_0=(p_0(x))^{-1}{\hat H}_0p_0(x)=-(\d_n-\fr{1}{2}\bb\phi_n(x))(\d_n+\fr{1}{2}\bb\phi_n(x))~~.
\label{ham2}
\ee
The two operators ${\hat H}_0$ and $H_0$ have the same eigenvalues but the latter has
the additional property that its left and right eigenfunctions are identical.
We note that 
\be
H_0=-\d_n^2+\fr{1}{4}\bb^2(\phi_n(x))^2-\fr{1}{2}\bb\phi_{nn}~~, \label{ham3}
\ee
and finally that eq(\ref{ann2}) implies that
\be
H_0p_0(x)=0~~.
\label{ham4}
\ee

\section{\label{slave}Slave Equations}

For a number of reasons both mathematical and physical it interesting
to examine the effect on the motion of the particle of small
external changes. In this paper we consider two types of change, namely,
a change of initial conditions where $\{y_n\}\rightarrow\{y_n+\dd y_n\}$, and
the application of an external field where the drift velocity is modified by making the replacement
$\phi_n(X)\rightarrow\phi_n(X)-J_af_{na}(X)$. The functions $\{f_{na}(X)\}$ are a basis set for the 
external influences in which we are interested.  A significant case is one where $f_{na}(X)=\d_nf_a(X)$.
This is equivalent to the replacement$\phi(X)\rightarrow \phi(X)-J_af_a(X)$.
The choice $f_a(X)=X_a~,~~a=1,2,3$ describes the action of an {\it electric field} on the motion of the 
particle. We will be interested also in situations in which $f_{na}(X)$ is not the gradient 
of a scalar field. An external {\it magnetic field} falls into this category. 

We assume that all the effects are weak so it is sufficient to 
consider a limit in which for a given noise function the induced change, 
$X_n\rightarrow X_n+\dd X_n$~, is such that $\dd X_n=O(\dd y)$ or $O(J)$. 

\subsection{\label{varip}Variation of Initial Position}

The effect of an infinitesimal variation of the initial position of the particle can be
computed from the behaviour of the quantities $\{\Xi_{mn}\}$ where
\be
\Xi_{mn}=\fr{\d X_m}{\d y_n}~~.
\ee
The differentiation is performed for a fixed sample $\{W_n(t)\}$ of the
noise. These variables satisfy the equation
\be
{\dot{\Xi}_{mn}}=-\phi_{mk}(X)\Xi_{kn}~~.
\label{slave1}
\ee
We refer to the $\{\Xi_{mn}\}$ as slave variables and to eq(\ref{slave1}) as
a slave equation. The slaves depend on the noise only through the 
original position variables $\{X_m\}$ but the latter are not influenced
by the evolution of the slaves. These variables were introduced by Graham \cite{GRAHAM}
who first elucidated their relationship with the underlying supersymmetric 
structure of the theory. He identified the relationship of the associated
decay exponent with a supersymmetric eigenvalue problem. 

Because it is a useful introduction to the properties of slave variables we 
will briefly review in our notation some of his results. The first step
is to introduce the joint probability distribution $P(x,\xi,t)$ where
\be
P(x,\xi,t)=\la\dd(x-X(t))\dd(\xi-\Xi(t))\ra~~.
\ee
It is standard theory,  given eq(\ref{diff1}) and eq(\ref{slave1}),
that
\be
\fr{\d}{\d t}P(x,\xi,t)=T\d_n(\d_n+\bb\phi_n)P(x,\xi,t)
                  +\fr{\d}{\d\xi_{mn}}\phi_{mk}(x)\xi_{kn}P(x,\xi,t)~~.
\label{diff3}
\ee

We introduce the quantities $P_{mn}(x)$, where
\be
P_{mn}(x,t)=\int d\xi \xi_{mn}P(x,\xi,t)~~.
\ee
At time $t$ 
\be
\la \Xi_{mn}\ra=\xib_{mn}=\int dxd\xi \xi_{mn}P(x,\xi,t)=\int dxP_{mn}(x,t)~~.
\ee
Here $d\xi$ is the volume element in $\xi$-space.
From eq(\ref{diff3}) we can show that
\be
\fr{\d}{\d t}P_{mn}(x,t)=T\d_n(\d_n+\bb\phi_n)P_{mn}(x,t)
                  -\phi_{mk}(x)P_{kn}(x,t)~~.
\label{diff4}
\ee

Following Graham \cite{GRAHAM}, we introduce a set, $\{a_n\}$, of anticommuting variables,
one for each of the variables $\{x_n\}$, together with their hermitian conjugates, $\{a^\dag_n\}$,
and a fermionic ground state $|0\ra$~. The new fermionic variables satisfy the (Clifford)
algebra
\be
\{a_m,a_n\}=\{a^\dag_m,a^\dag_n\}=0~~,
\ee
and
\be
\{a_m,a^\dag_n\}=\dd_{mn}~~.
\ee
The fermionic ground state satisfies
\be
a_m|0\ra=0~~,
\ee
and is normalised so that $\la 0|0\ra=1$~.

We now augment the operator $H_0$ by defining $H$ where
\be
H=H_0+\bb\phi_{mn}(x)a^\dag_ma_n~~,
\label{ham5}
\ee
and define the state $\psi_0$ so that
\be
\psi_0=p_0(x)|0\ra~~.
\ee
It is easily checked that
\be
H\psi_0=0~~.
\ee
We now introduce $F_{mn}(x,t)$ where
\be
P_{mn}(x,t)=F_{mn}(x,t)P_0(x)~~,
\ee
and define the state
\be
\psi_n(t)=F_{mn}(x,t)a^\dag_m\psi_0~~.
\ee
Eq(\ref{diff4}) can now be put in the form
\be
\fr{\d}{\d t}\psi_n(t)=-TH\psi_n(t)~~.
\label{diff4a}
\ee
The formal solution to this equation is
\be
\psi_n(t)=e^{-THt}\psi_n(0)~~.
\label{diff4b}
\ee
It follows that the long time behaviour of $F_{mn}(x,t)$ and hence
that of $P_{mn}(x,t)$ and finally of $\xib_{mn}$ is dominated by
the lowest eigenvalue, $E$, of $H$ in the one fermion sector of the state space.
Typically we expect that at large times
\be
\xib_{mn}\sim e^{-TE t}~~.
\ee
Therefore the  decay exponent $\nu$, that governs the large time statistical 
evolution of line elements is given by
\be
\nu=TE~~.
\ee
This is the essence of Graham's
results \cite{GRAHAM}~. We will examine later the supersymmetric structure
of $H$. 

The above approach can be extended to area elements. We consider the 
infinitesimal area element at $t=0$ which has the form $dy_m\wg dy_n$.
At a later time $t$ it has evolved into the element
\be
dX_m\wg dX_n=\Xi_{mk}\Xi_{nl}dy_k\wg dy_l~~.
\ee
That is
\be
dX_m\wg dX_n=\Xi_{mn,kl}dy_k\wg dy_l~~,
\ee
where
\be
\Xi_{mn,kl}=\fr{1}{2}(\Xi_{mk}\Xi_{nl}-\Xi_{nk}\Xi_{ml})~~.
\ee
Clearly 
\be
\la \Xi_{mn,kl}\ra=\xib_{mn,kl}=\int dxd\xi\xi_{mn,kl}P(x,\xi,t)~~.
\ee
where
\be
\xi_{mn,kl}=\fr{1}{2}(\xi_{mk}\xi_{nl}-\xi_{ml}\xi_{nk})~~.
\ee
We define $P_{mn,kl}(x,t)$ and $F_{mn,kl}(x,t)$ so that
\be
P_{mn,kl}(x,t)=F_{mn,kl}(x,t)P_0(x)=\int d\xi\xi_{mn,kl}P(x,\xi,t)~~.
\ee
From eq(\ref{diff3}) we can show that
$$
\fr{\d}{\d t}P_{mn,kl}(x,t)=T\d_n(\d_n+\bb\phi_n(x))P_{mn,kl}(x,t)
~~~~~~~~~~~~~~~~~~~~~~~~~~~~~~~~~~~~~
$$
\be
~~~~~~~~~~~~~~~~~~~~~~~~~~~~~~~~~~~~~~~
                   -\phi_{mk'}(x)P_{k'n,kl}(x,t)-\phi_{nk'}(x)P_{mk',kl}(x,t)~~.
\label{diff5}
\ee
We define
\be
\psi_{kl}=F_{mn,kl}(x,t)a^\dag_ma^\dag_n\psi_0~~.
\ee
Eq(\ref{diff5}) is equivalent to the result
\be
\fr{\d}{\d t}\psi_{kl}=-TH\psi_{kl}~~.
\label{diff5a}
\ee
It follows that the behaviour of $\psi_{kl}$ at large times, and hence that of
$P_{mn,kl}(x,t)$ and $\la \Xi_{mn,kl}\ra$, is dominated by the lowest
eigenvalue of $H$ in the two-fermion sector. This is again one
of Graham's results \cite{GRAHAM}.

Clearly the argument can be generalised to higher dimensional volume
elements. We will not exhibit the argument in detail. The important point is
that the time evolution of a  state, $\psi$, at any fermionic level is governed by the 
equation
\be
\fr{\d}{\d t}\psi=-TH\psi~~.
\label{diff5b}
\ee
Decay exponents for differential volume elements of
any dimension can be calculated through the eigenvalues of the supersymmetric
Hamiltonian.

\subsection{\label{extfld}External Fields}

The effect of an external field can be induced by 
replacing eq(\ref{diff1}) with
\be
{\dot{X}}_n=-\phi_n(X)+J_af_{na}(X)+W_n(t)~~.
\label{diff1b}
\ee
Of particular interest is the {\it electric} case where $f_{na}(X)=\d_nf_a(X)$. 
We can think of the set $\{f_a(X)\}$ as a set of electric potential functions.
If we make the further specialisation 
\be
f_a(X)=X_a~~,~~(a=1,2,3)~~,
\ee
the resulting field is an uniform electric field $(J_1,J_2,J_3)$~.
The {\it magnetic} case is one in which $\d_mf_{na}(X)-\d_nf_{ma}(X)\ne 0$ and $f_{na}(X)$ 
is not the derivative of a function $\{f_a(X)\}$. We will consider the magnetic case in detail later.
The following derivation covers both cases.

If we solve eq(\ref{diff1b}) for a particular sample of the noise process
with a given initial condition, $X(0)=0$, say, then at a later time, $t$, 
the value of $X$ will depend on the value of $J_n$ and we can regard the 
solution at this time as providing a map from $J$-space to $X$-space.
Therefore curves and surfaces in $J$-space will have images in $X$-space
that are also curves and surfaces. We can examine the differential version 
of this map by introducing appropriate slave variables $\Xi_{mn}$ given by
\be
\Xi_{ma}=\fr{\d X_m}{\d J_a}~~.
\ee
we have then
\be
dX_m=\Xi_{ma}dJ_a~~.
\label{diffntl1}
\ee
By differentiation eq(\ref{diff1b}) with respect to $J_a$
we can obtain an equation for the evolution of $\Xi$ namely
\be
\dot{\Xi}_{ma}=-\phi_{mk}(X)\Xi_{ka}+f_{ma}(X)~~,
\label{slave2}
\ee
Of course we have set $J_a=0$ after differentiation.
The initial condition for the slave is $\Xi_{ma}(0)=0$
since we assume that $X_n(0)$ is independent of $J_a$~. 

Eq(\ref{diffntl1}) implies that
\be
d\la X_m\ra=\la dX_m\ra=\la\Xi_{ma}\ra dJ_a~~.
\ee
that is
\be
\fr{\d}{\d J_a}\la X_m\ra=\la\Xi_{ma}\ra~~.
\ee
Thus $\la\Xi_{ma}\ra$ can be understood as a set of susceptibilities of the
system at time $t$ with respect to the various contributions to the external field 
in the weak field limit $J_a\rightarrow 0$~. 

Eq(\ref{diff1}) and eq(\ref{slave2}) imply that the joint probability distribution,
$P(x,\xi,t)$, given by
\be
P(x,\xi,t)=\la\dd(x-X(t))\dd(\xi-\Xi(t))\ra
\ee satisfies
$$
\fr{\d}{\d t}P(x,\xi,t)=T\d_n(\d_n+\bb\phi_n(x))P(x,\xi,t)
~~~~~~~~~~~~~~~~~~~~~~~~~~~~~~~~~~~~~~~~~~~
$$
\be
~~~~~~~~~~~~~~~~~~~~~~~~~~~~~~~~~~~~~~~~~~
                       +\fr{\d}{\d \xi_{ma}}(\phi_{mk}\xi_{ka}(x)-f_{ma}(x))P(x,\xi,t)~~.
\label{diff6}
\ee
An important distinction between this case and the previous one is that we expect
the joint probability distribution to approach a static form at large time.
That is $P(x,\xi,t)\rightarrow P(x,\xi)$ as $t\rightarrow \infty$, where
\be
\d_n(\d_n+\bb\phi_n(x))P(x,\xi)
    +\bb\fr{\d}{\d \xi_{ma}}(\phi_{mk}(x)\xi_{ka}-f_{ma}(x))P(x,\xi)=0~~.
\label{diff7}
\ee
We shall be mainly interested in this asymptotic static situation from now on.
Following the pattern established in subsection \ref{varip} we define, for this new case, 
$P_{ma}(x)$ and $F_{ma}(x)$ so that
\be
P_{ma}(x)=F_{ma}(x)P_0(x)=\int d\xi\xi_{ma}P(x,\xi)~~.
\ee
Hence
\be
\la\Xi_{ma}\ra=\xib_{ma}=\int dxF_{ma}(x)P_0(x)~~.
\ee
We can interpret $F_{ma}(x)$ as the contribution to the susceptibility
from the particles in the neighbourhood of the point $x$.
From eq(\ref{diff7}) we can deduce that
\be
\d_k(\d_k+\bb\phi_k(x))P_{ma}(x)
    -\bb\phi_{mk}(x)P_{ka}(x)+\bb f_{ma}(x)P_0(x)=0~~.
\label{diff8}
\ee
Invoking the fermionic variables introduced above and setting 
\be
\psi_a=F_{ma}(x)a^\dag_m\psi_0 ~~,
\label{locsus}
\ee
we find that eq(\ref{diff8}) can be expressed in the form
\be
H\psi_a-\bb f_{ma}(x)a^\dag_m\psi_0=0~~.
\label{diff8a}
\ee

\subsubsection{\label{arelem}Area Elements}

The discussion can be extended to area elements. At time $t$ the solutions of
eq(\ref{diff1b}) map the $J$-space area element $dJ_a\wg dJ_b$ into
the $X$-space area element $dX_m\wg dX_n$ thus
\be
dX_m\wg dX_n=\Xi_{ma}\Xi_{nb}dJ_a\wg dJ_b=\Xi_{mn,ab}dJ_a\wg dJ_b~~,
\ee
where
\be
\Xi_{mn,ab}=\fr{1}{2}(\Xi_{ma}\Xi_{nb}-\Xi_{na}\Xi_{mb})~~.
\ee
We can think of $\la\Xi_{mn,ab}\ra$ at large $t$ as a generalised area susceptibility.
The relationship of this susceptibility for the supersymmetric structure can be
understood by defining $P_{mn,ab}(x)$ and $F_{mn,ab}(x)$ so that
\be
P_{mn,ab}(x)=F_{mn,ab}(x)P_0(x)=\int d\xi \xi_{mn,ab}P(x,\xi)~~,
\ee
where
\be
\xi_{mn,ab}=\fr{1}{2}(\xi_{ma}\xi_{nb}-\xi_{na}\xi_{mb})~~.
\ee
Hence
\be
\la\Xi_{mn,ab}\ra=\xib_{mn,ab}=\int dxF_{mn,ab}(x)P_0(x)~~.
\ee
From eq(\ref{diff7}) we can show that
$$
\d_p(\d_p+\bb\phi_p(x))P_{mn,ab}(x)
~~~~~~~~~~~~~~~~~~~~~~~~~~~~~~~~~~~~~~~~~~~~~~~~~~~~~~~~~~~~~~~~
$$
$$
~~~~~~~~~~~~
            -\bb\phi_{mm'}P_{m'n,ab}(x)-\bb\phi_{nn'}(x)P_{mn',ab}(x)
~~~~~~~~~~~~~~~~~~~~~~~~~~~~~~~~~~~~~~~~~~
$$
\be
~~~~~~~~~~~~~~~~~
+\bb(f_{ma}(x)P_{nb}(x)+f_{nb}(x)P_{ma}(x)-f_{mb}(x)P_{na}(x)
                                 -f_{na}(x)P_{mb}(x))=0~~.
\label{diff9}
\ee
If we now define
\be
\psi_{ab}=F_{mn,ab}(x)a^\dag_ma^\dag_n\psi_0~~,
\ee
then eq(\ref{diff9}) can be put in the form
\be
H\psi_{ab}-\bb(f_{ma}(x)a^\dag_m\psi_b-f_{nb}(x)a^\dag_n\psi_a)=0~~.
\label{diff9a}
\ee
From eq(\ref{diff9a}) we can calculate the area susceptibility
in terms of the eigenstates of $H$ at the second fermionic level.
The procedure can be generalised to deal with volume elements (of
any dimension). Higher order susceptibilities of this kind fit
naturally into the supersymmetric approach to the diffusion model.
However, while we feel that they are an interesting area for study, 
in contrast to the standard susceptibility to the action of an
external field, their observability is unclear.

\section{\label{simul}Simulation}

We will test some of the theoretical predictions of the stochastic 
model by simulation. The first order algorithm corresponding to eq(\ref{diff1})
with correlator given by eq(\ref{stoch}) is a process in which for a time step
$\dd t$, $X\rightarrow X+\dd X$
where
\be
\dd X_n=-\phi_n(X)~\dd t+\sqrt{2T\dd t}~\eta_n~~.
\ee
Here $\eta_n$ is a Gaussian random variable with correlator
\be
\la \eta_m\eta_n\ra=\dd_{mn}~~.
\ee
The errors in the simulated probability distributions are $O(\dd t)$~.
In order to control errors more easily we use a second order algorithm
with an intermediate step $X\rightarrow X^{(1)}=X+\dd X^{(1)}$ where
\be
\dd X_n^{(1)}=-\fr{1}{2}\phi_n(X)~\dd t+\sqrt{T\dd t}~\eta^{(1)}_n~~.
\ee
The final step is
\be
\dd X_n=-\phi_n(X^{(1)})~\dd t+\sqrt{T\dd t}~(\eta_n^{(1)}+\eta_n^{(2)})~~'
\ee
where $\eta^{(1)}_n$ and $\eta^{(2)}_n$ are Gaussian random variables
of the same type as $\eta$ above. This procedure, which is a stochastic
second order Runge-Kutta technique \cite{} can be extended to
include the slave equation in an obvious way. The resulting probability
distributions have errors that are $O(\dd t^2)$~.

\section{\label{ssform}Supersymmetric Formulation}

Apart from the straightforward case in which $\phi(X)$ is a
quadratic function, the equations for $X$ and $\Xi$ are 
intrinsically non-linear. It is a remarkable and significant fact that
associated with such non-linear processes is an underlying
linear algebraic structure 
\cite{WITTEN,KURCHAN1,KURCHAN2,KURCHAN3,GRAHAM,NICOLAI1,NICOLAI2}. 
The supersymmetric formulation of the theory is a powerful way of organising 
this linear algebraic structure. Although well known, see in particular 
\cite{KURCHAN1,KURCHAN2}, in the interests of completeness 
and to establish some notation we give an account of it here.

The supersymmetric structure is made evident by the construction of the
operators
\be
Q=-ia_l\left(\d_l-\fr{1}{2}\bb\phi_l\right)~~,
\ee
and
\be
Q^\dag=-ia^\dag_l\left(\d_l+\fr{1}{2}\bb\phi_l\right)~~.
\ee
They satisfy
\be
Q^2=(Q^\dag)^2=0~~.
\ee
The combinations $Q_1=Q+Q^\dag$ and $Q_2=i(Q-Q^\dag)$ are hermitian
and satisfy
\be
Q_1Q_2+Q_2Q_1=0~~.
\ee
We also have
\be
Q_1^2=Q_2^2=H~~,
\label{susy}
\ee
where 
\be
H=QQ^\dag+Q^\dag Q~~,
\label{susy1}
\ee
and $H$ is given by eq(\ref{ham5}).

\subsection{The Kernel of $H$}

From eq(\ref{ham5}) we see that $\psi_0$ is in the kernel of $H$~. 
Eq(\ref{susy}) then tells us that
$\psi_0$ is in the kernel of $Q_1$ and that of $Q_2$ also. In fact any eigenstate in
the kernel of $H$ is annihilated by both $Q_1$ and $Q_2$ and hence by both $Q$ 
and $Q^\dag$~.

The operator $\LL=a^\dag_ka_k$ counts the number of $a$-excitations in a state.
It commutes with $H$ and can be used to classify the eigenstates of H. In
particular the basis states in the kernel of $H$ can be chosen so that they 
are each associated with an eigenvalue of $\LL$, that in our three-dimensional model can
take one of the values 0, 1, 2, or 3. If there is only one basis state in
the kernel of $H$ then it can be taken to be $\psi_0$ and
\be
\LL\psi_0=0~~.
\ee
In fact because our model involves only a simple flat euclidean space, 
the kernel of $H$ is one dimensional and spanned by $\psi_0$ \cite{WITTEN}. 

\subsection{Eigensubspace structure of $H$}
We also see that
if $\chi$ is an eigenstate of $H$ with eigenvalue $E\ne 0$ then so is the
orthogonal state $\chi'=Q_1\chi$ (or equivalently $Q_2\chi$). That is, 
the eigenstates of $H$ that are not in its kernel come in pairs with equal 
eigenvalues. There is no corresponding constraint on the eigenstates lying 
within the kernel. 

On any subspace $\Hc_{\ll}$ with $\LL=\ll$ and $0\le\ll\le 3$, that does not 
intersect the kernel of $H$ we have the projection operators
\be
\Om^{(-)}=Q^\dag Q\fr{1}{H}~~,
\ee
and 
\be
\Om^{(+)}=QQ^\dag\fr{1}{H}~~.
\ee
They satisfy 
\be
\Om^{(+)}+\Om^{(-)}=1~~,
\ee
and other appropriate relations. It follows that, in general, $\Hc_{\ll}$ is a
direct sum of two orthogonal subspaces $\Hc_{\ll}^{(\pm)}=\Om^{(\pm)}\Hc_{\ll}$
that satisfy
\be
\Om^{(\pm)}\Hc_{\ll}^{(\pm)}=\Hc_{\ll}^{(\pm)}~~,~~~~~~~~~~~~
       \mbox{and}~~~~~~~~~~~~\Om^{(\pm)}\Hc_{\ll}^{(\mp)}=0~~.
\ee
Therefore the basis states of $\Hc_\ll$ can be chosen so that they lie in 
either $\Hc^{(+)}_\ll$ or $\Hc^{(-)}_\ll$~. In an obvious notation we have
\be
H\chi^{(\pm)}_{\ll\aa}=E^{(\pm)}_{\ll\aa}\chi^{(\pm)}_{\ll\aa}~~.
\ee 
It is then obvious from the discussion that we could choose phases so that
\be
\chi^{(-)}_{\ll\aa}=\fr{1}{\sqrt{E^{(+)}_{(\ll-1)\aa}}}Q^\dag \chi^{(+)}_{(\ll-1)\aa}~~,
\ee
with $E^{(-)}_{\ll\aa}=E^{(+)}_{(\ll-1)\aa}$ and
\be
\chi^{(+)}_{\ll\aa}=\fr{1}{\sqrt{E^{(-)}_{\ll\aa}}}Q\chi^{(-)}_{(\ll+1)\aa}~~,
\ee
and $E^{(+)}_{\ll\aa}=E^{(-)}_{(\ll+1)\aa}$~. 

It follows then that any state in $\Hc_{\ll}$ of the form $Q^\dag\chi$ lies 
in $\Hc_{\ll}^{(-)}$~.  It is also true that any state that lies in $\Hc^{(-)}_{\ll}$
can be put in the form $Q^\dag\chi$ where $\chi$ lies in $\Hc_{(\ll-1)}$~. The state
$\chi$ is not unique but the ambiguity can be removed by requiring that 
$\chi$ lies in $\Hc^{(+)}_{(\ll-1)}$~. That is, by imposing the condition $Q\chi=0$~.
Similarly a state that lies in $\Hc^{(+)}_{\ll}$ can be put in the form $Q\rho$
for some state $\rho$ that lies in $\Hc^{(-)}_{(\ll+1)}$ and that satisfies $Q^\dag\rho=0$~.

The above analysis has to be modified for the extreme subspaces $\Hc_0$ and 
$\Hc_3$~~. The subspace $\Hc_0$ is a direct sum of $\Hc_0^{(+)}$ and the 
one dimensional subspace containing $\psi_0$~. The subspace $\Hc_3$ is
identical with $\Hc_3^{(-)}$~.

\section{\label{emaganal}Electromagnetic Analogy}

An investigation of the eigenstates and eigenvalues of the supersymmetric Hamiltonian
is equivalent to studying the evolution equation, eq(\ref{diff5b}) at each fermionic level.
It is helpful, then, to think in terms of an analogy with electromagnetic systems. 
The analogy makes it natural to introduce the Principle of Minimum Dissipation that
we establish in section \ref{mindiss}. In turn this will help us to establish in a 
natural way the structure of the eigenstates of the supersymmetric Hamiltonian and
in particular, in the low temperature limit as elucidated by T\v anase-Nicola and Kurchan \cite{KURCHAN1} 

\subsection{Zero Fermion Level}

Consider, for example, a state $\psi$ in the zero fermion sector. It has the form
\be
\psi=V(x)\psi_0~~.
\label{typical}
\ee
In the context of our proposed electromagnetic analogy, it is convenient to associate 
with it a charge density distribution $\rho(x)$ where
\be
\rho(x)=V(x)P_0(x)~~.
\label{newdiff1a}
\ee
(Of course $\rho$ and $V$ may also have a dependence on $t$ that we leave implicit).

The image of $\psi$ in the one fermion sector is $\chi=iQ^\dag\psi$~. It
follows immediately that $Q^\dag\chi=0$ and therefore that $\chi\in\Hc_1^{(-)}$~.
Evaluating $\chi$ explicitly we find
\be
\chi=a^\dag_n(\d_n+\fr{1}{2}\bb\phi_n(x))V(x)\psi_0=(\d_nV(x))a^\dag_n\psi_0~~.
\ee
We continue the electromagnetic analogy by associating with $\chi$ an electric current density 
$\bj$ where
\be
\bj=-TP_0(x)\nabla V(x)~~.
\label{newdiff1b}
\ee
The justification for these identifications is that, if we assume that
$\psi$ satisfies eq(\ref{diff5b}), we find that
\be
\fr{\d}{\d t}\rho=-\nabla\cdot\bj~~,
\label{newdiff1}
\ee
which is indeed the law of conservation of charge. We can complete the analogy by identifying
$V(x)$ as the voltage distribution and $-\nabla V(x)$ as the electric field acting on the
charge. The static probability density distribution $P_0(x)$ plays a dual role.
It is the capacity per unit volume and also determines the local conductivity, which is $TP_0(x)$~.

The eigenvalue equation for the supersymmetric Hamiltonian in the zero fermion sector
can be re-expressed in the analogy by making the replacements $V=V^{(0)}e^{-TEt}$,
$\rho=\rho^{(0)}e^{-TEt}$ and $\bj\rightarrow \bj^{(0)}e^{-TEt}$, where the 
superfix $(0)$ indicates that the quantity is evaluated at $t=0$ and $E$ is the eigenvalue
of the supersymmetric Hamiltonian. We have then a new form for the eigenvalue equation
from eq(\ref{newdiff1}) 
\be
\nabla\cdot\bj=TE\rho~~.
\label{eig1}
\ee

We obtain in the standard way the integral version of eq(\ref{newdiff1}) namely
\be
\int_R dx \fr{\d}{\d t}\rho=\int_Sd\Sb\cdot\bj~~,
\label{integ1}
\ee
where $R$ is an arbitrary region of space and $S$ is its bounding surface.

The norm of the state $\psi$ also has a natural interpretation. If we define $U$ so that
\be
U=\fr{1}{2}(\psi,\psi)~~,
\ee
then we find
\be
U=\fr{1}{2}\int dx\rho(x)V(x)~~.
\ee
That is, the norm of $\psi$ has can be interpreted as the electrostatic energy of the state.

\subsection{\label{OF1}One Fermion Level}

A parallel analysis can be carried out at fermionic level one. A state $\psi\in \Hc^{(+)}_1$
has the form
\be
\psi=H_n(x)a_n^\dag\psi_0~~,
\ee
and is annihilated by $Q$~. That is
\be
iQ\psi=(\d_m-\fr{1}{2}\bb\phi_m(x))\psi_0=[(\d_m-\bb\phi_m(x))H_m(x)]\psi_0=0~.
\ee
It follows that
\be
\d_mP_0(x)H_m(x)=0~~.
\label{div0}
\ee
That is, the state is associated with a conserved vector field, $B_n(x)$,
where
\be
B_n(x)=P_0(x)H_n(x)~~.
\label{newdiff2a}
\ee 
If we adopt eq(\ref{diff5b}) as the equation of motion,
then we have
\be
\fr{\d}{\d t}P_0(x)H_n=T\d_m[P_0(x)(\d_mH_n-\d_nH_m)]~~.
\label{newdiff2aa}
\ee

We can interpret this equation in terms of an electromagnetic analogy in the following way.
We identify $\Hb=(H_1,H_2,H_3)$ as the magnetic field and $\Bb=(B_1,B_2,B_3)$
as the magnetic induction field. Eq(\ref{newdiff2a}) then implies that $\Bb=P_0(x)\Hb$ and
hence that $P_0(x)$ plays the role of (position dependent) magnetic permeability. 
We introduce an electric current $\bj$ and an electric field $\Eb$. We link them through
a space dependent version of Ohm's Law,
\be
\bj=\ss(x)\Eb~~,
\label{Ohm1}
\ee
where the space dependent conductivity $\ss(x)$ is given by
\be
\ss(x)=\fr{\bb}{P_0(x)}~~.
\label{Ohm2}
\ee
Eq(\ref{div0}) is easily restated as
\be
\nabla\cdot\Bb=0~~.
\label{div00}
\ee
Eq(\ref{newdiff2aa}) is reproduced by combining Faraday's law of induction
\be
\fr{\d}{\d t}\Bb=-\nabla\times\Eb~~,
\label{Faraday}
\ee
with Ohm's law and Amp$\grave{\mbox{e}}$re's law
\be
\nabla\times\Hb=\bj~~.
\label{Ampere}
\ee
That is we have
\be 
\fr{\d}{\d t}\Bb=-T\nabla\times(P_0(x)\nabla\times\Hb)~~.
\label{newdiff2b}
\ee

By setting $\Hb=\Hb^{(0)}e^{-TEt}$ {\it etc.} where $E$ is an eigenvalue of the supersymmetric
Hamiltonian at the one fermion level, we obtain the eigenvalue equation in the form
\be
\nabla\times(P_0(x)\nabla\times\Hb^{(0)})=EP_0(x)\Hb^{(0)}~~.
\ee
It is interesting that allowing for the position dependent magnetic susceptibility
represented by $P_0(x)$ this equation has a form analogous to the London Equation
for a superconducting medium \cite{LON,DEGN}.

We can obtain an integral version of eq(\ref{Faraday}) in the form
\be
\fr{\d}{\d t}\int_S \Bb\cdot d\Sb=-\int_C \Eb\cdot d\bx~~,
\label{integ2}
\ee
where $S$ is an arbitrary surface and $C$ is its bounding loop with the conventional
right-handed orientation.  We also have the integral version of eq(\ref{Ampere}) namely
\be
\int_S \bj\cdot d\Sb=\int_C \Hb\cdot d\bx~~.
\label{integ3}
\ee

Again, from the norm of the state, we can identify an energy $U$ given by
\be
U=\fr{1}{2}(\psi,\psi)= \fr{1}{2}\int dx \Hb\cdot\Bb~~.
\ee
In the context of the electromagnetic analogy this a very natural result.

\subsection{Two and Three Fermion Level}

For completeness we discuss briefly the two and three fermion sectors.

The evolution of the states in the subspace $\Hc^{(-)}_2$ can be obtained by the
application of $Q^\dag$ to the states in $\Hc^{(+)}_1$ that we have just discussed.

The evolution of the states in $\Hc^{(+)}_2$ can most easily be investigated by 
applying $Q$ to the corresponding states in $\Hc^{(-)}_3$. These states, because
we are in three dimensions, have the form
\be
\psi=M(x)\fr{1}{6}\eps_{lmn}a^\dag_la^\dag_ma^\dag_n\psi_0~~.
\ee
The supersymmetric equation of motion then implies
\be
\fr{\d}{\d t}M=T\d_l(\d_l-\bb\phi_l)M~~.
\ee
This is very close to the original diffusion equation for $P(x,t)$ but with the
replacement $\phi\rightarrow -\phi$ \cite{JUNKER}. However since $\exp\{\bb\phi(x)\}$
is not a normalisable wavefunction, there is no stationary solution 
corresponding to a zero eigenvalue of the Hamiltonian $H$~.

\section{\label{mindiss}Principle of Minimum Dissipation}

We derive here the Principle of Minimum Dissipation in the context of
the electromagnetic analogy. We will be able to make use of the principle 
to derive the spatial structure of wavefunctions with low lying eigenvalues 
in the large $\bb$ limit.

\subsection{Zero Fermion Sector}

Eq(\ref{newdiff1}) implies that the energy dissipation rate $W$ is
\be
W=-\fr{\d U}{\d t}=\int dxV\d_nj_n=-\int dx(\d_nVj_n)~~.
\ee
The last step is achieved by integrating by parts. The result is clearly consistent
with the idea of an electric current being driven by by a potential gradient.
It follows that
\be
W=\bb\int dx\fr{j_n^2}{P_0(x)}~~.
\label{dissip1}
\ee

It is a well known principle that, for a given set of sources, the true current 
distribution minimises the energy dissipation as calculated from eq(\ref{dissip1}). 
We can verify this in the present context by noting that if we modify $j_n$ so that
\be
j_n=-TP_0(x)\d_nV+\d_mA_{mn}~~,
\ee
where $A_{mn}$ is an arbitrary antisymmetric tensor field then $\d_nj_n$ is unchanged
and therefore $j_n$ retains the same (time-dependent) source distribution as the original current distribution. 
However the new dissipation rate calculated from eq(\ref{dissip1}) is
\be
W=\int dx TP_0(x)(\d_nV)^2+\int dx\fr{1}{P_0(x)}(\d_mA_{mn})^2~~.
\ee
The cross term vanishes after integration by parts. Clearly the minimum
of $W$ occurs when $\d_mA_{mn}$ vanishes and $j_n$ acquires its physical value.
This outcome can be described by saying that for given sources the true current follows the path 
of least resistance. Under appropriate circumstances we can use the principle to
identify (or guess) the true current distribution given the source distribution. 
For example we will be interested later in circumstances in which $\bb$ is very large.
In this limiting regime the hills and valleys of $\phi(x)$ are highly magnified in the
rise and fall of the local resistivity, $(P_0(x))^{-1}$~. Indeed certain regions where $\phi(x)$
is large become effectively insulators. We are therefore in a good position 
to identify the complementary regions in which the current can flow easily and in which
in which $j_n^2$ remains substantial. The structure of these regions depends on the 
detailed shape of $\phi(x)$ but in the models we consider they will be filaments along lines of 
steepest descent. The calculation becomes essentially one in electric circuit theory
where the ideas of minimal dissipation are very familiar.

Of course approximation schemes based on the principle of minimal dissipation could be formulated
to deal with circumstances in which the temperature is well above the low temperature limit. We
do not pursue such calculations in this paper.

\subsection{One Fermion Sector}

Equivalent results can be obtained at the one fermion level.
From eq(\ref{newdiff2b}) we can verify that the dissipation $W$ is given by
\be
W=-\fr{\d U}{\d t}=\int dxP_0(x)(\nabla\times\Hb)^2=\bb\int dx\fr{1}{P_0(x)}\Eb^2=\int dx\bj\cdot \Eb~~.
\label{dissip2}
\ee
We can modify $\Eb$ by making the replacement $\Eb\rightarrow\Eb^{\mbox{new}}=\Eb+\nabla A$ where $A$ is a scalar field.
This leaves the law of induction unchanged. The dissipation however is changed to
\be
W=\bb\int dx\fr{1}{P_0(x)}(\Eb^{\mbox{new}})^2=\bb\int dx\fr{1}{P_0(x)}\Eb^2+\bb\int dx\fr{1}{P_0(x)}(\nabla A)^2.
\ee
It is clear that the minimum occurs when $\nabla A$ vanishes and $\Eb$ 
acquires its true value. We can see from this principle that for a given set
of sources the electric field will be greatest where $P_0(x)$ is
greatest and least where it is least. In the low temperature limit these variations are exaggerated
and this allows us to separate the space into regions of essentially zero electric field 
and complementary regions, confined to thin sheets in our model, where $\Eb$ is large.

Of course these results can be generalised to any fermionic level
but we shall not present them explicitly. They will help us in understanding the 
structure of the eigenstates of the supersymmetric Hamiltonian and their associated 
current distributions.

\section{\label{low_T}High-$\bb$ Limit and Low Lying Eigenvalues}

Witten \cite{WITTEN} emphasised the importance of the limit of high $\bb$ in analysing
the structure of stationary points of $\phi(x)$~. T{\v a}nase-Nicola and Kurchan 
\cite{KURCHAN1,KURCHAN2} interpreted the limit in the context of diffusion
processes and developed computational algorithms for exploring the neighbourhoods 
of stationary points. The model they considered was one in which $\phi(x)$ is a Morse function
and therefore has stationary points with non-degenerate quadratic structure.
We discuss the model here and show how we can develop the electromagnetic analogy 
to compute  decay exponents in the high $\bb$-limit. 

\subsection{\label{Zero} Zero Fermion Level}
We will investigate a model in which $\phi(x)$ is a Morse function with a set of $R$ minima
at the (sufficiently well separated) points $x=\xb^{(r)}$ where $r=1,\ldots,R$~. 
The minima have a quadratic structure  with the result that
in the neighbourhood of the minimum at $\xb^{(r)}$ we have
\be
\phi(x)=B_r+\fr{1}{2}(x_m-\xb_m^{(r)})A^{(r)}_{mn}(x_n-\xb_n^{(r)})~~,
\ee
where $B_r=\phi(\xb^{(r)})$ and $A^{(r)}_{mn}=\phi_{mn}(\xb^{(r))}$~.
Clearly
\be
A^{(r)}_{mn}=\sum_\gg \ll^{(r)}_\gg c^{(r)}_{\gg m}c^{(r)}_{\gg n}~~,
\ee
where $\ll^{(r)}_\gg$ are the eigenvalues of $\phi_{mn}(\xb^{(r)})$ and $c^{(r)}_{\gg n}$
are the corresponding eigenvectors.
We are interested in a situation in which $\bb$ is very large (low temperature).
In these circumstances $\exp\{-\bb\phi(x)\}$ is
significantly non-zero only in the neighbourhoods of the minima. We have
\be
\exp\{-\bb\phi(x)\}\simeq
\sum_r \exp\{-\bb [B_r+\fr{1}{2}(x_m-\xb_m^{(r))})A^{(r)}_{mn}(x_n-\xb_n^{(r)})]\}~~.
\ee
This approximation is adequate for computing the limiting expectation values of appropriately
smooth and slowly varying functions of $x$~. However we note parenthetically that there
are circumstances
in which we may wish to compute the next-to-leading corrections and that this can be
achieved by including terms in the eigenfunction approximation that involve third derivatives
of $\phi(x)$~. The effect can be accounted for by shifting the mean of $x_m$ from $\xb_m^{(r)}$
by an amount that is $O(\bb^{-1})$~.

The partition function, in this approximation, can be expressed in the form
\be
Z=\sum_r e^{-\bb B_r}Z^{(r)}~~,
\label{PART1}
\ee
where
\be
Z^{(r)}=\int dx\exp\{-\bb\fr{1}{2}(x_m-\xb_m^{(r))})A^{(r)}_{mn}(x_n-\xb_n^{(r)})\}~~.
\label{PART2}
\ee
If we define $w_r$ so that
\be
w_r=\sqrt{e^{-\bb B_r}Z^{(r)}/Z}~~,
\label{PART3}
\ee
then we can interpret $w_r^2$ as the probability of the particle being
in the neighbourhood of the $r$th minimum. The stationary probability
distribution is approximately
\be
P_0(x)\simeq \sum_r w_r^2P_0^{(r)}(x)~~,
\ee
where $P_0^{(r)}(x)$ is the probability distribution appropriate to the $r$th minimum.
Similarly the null state of $H$ can be expressed as
\be
\psi_0(x)\simeq \sum_r w_r p_0^{(r)}(x)|0\ra~~,
\ee
where $p_0^{(r)}(x)=\sqrt{P_0^{(r)}(x)}$~.

In fact we can use the states $\psi^{(r)}=p_0^{(r)}(x)|0\ra$ as an essentially orthonormal basis 
for the low eigenvalue subspace of $\Hc_0$~. The eigenstates of $H$ in this subspace have the form
\be
\psi_\aa=v_\aa(x)p_0(x)|0\ra\simeq\sum_r v_\aa(\xb^{(r)})w_r\psi^{(r)}~~,
\ee
where $v_\aa(x)$ is slowly varying as a function of $x$ and $v_0(x)\equiv 1$~. Because of the
effective orthonormality of the eigenfunctions and of the basis set the
matrix with elements $v_\aa(\xb^{(r)})w_r$ is orthogonal,
that is
\be
\sum_rv_\aa(\xb^{(r)})w_rv_{\aa'}(\xb^{(r)})w_r=\dd_{\aa\aa'}~~,
\label{ORTH1}
\ee
and
\be
\sum_\aa v_\aa(\xb^{(r)})w_rv_\aa(\xb^{(r')})w_{r'}=\dd_{rr'}~~,
\label{ORTH2}
\ee

We can extend the basis for the subspace $\Hc_0$ in two ways. First we can add 
excited states associated with the quadratic structure at each minimum.
These states are, like the corresponding ground states $\psi^{(r)}$, peaked 
about each minimum. They are of course orthogonal to these ground states. The associated
eigenvalues are proportional to $\bb$~. Second we note that, in the limit of large $\bb$, all stationary points, 
including saddles and maxima, yield minima of the potential in the Hamiltonian $H_0$~.
We can include then the ground states and higher states associated with the
quadratic structure of each of the stationary points. These states, in the high-$\bb$
limit, are all orthogonal to one another and to the states associated with the minima
of $\phi(x)$ discussed above. Again all of the states have eigenvalues that are
proportional to $\bb$~. Of course we must cut off the eigenvalue expansions before
the corresponding eigenvalues rise to the point where we expect serious mixing between
the true eigenstates of $H_0$~. However if the quantities we wish to evaluate
are dominated by the lower lying of these large eigenvalue states this mixing will
not be significant.

We can account for the lowest of these eigenstates that are peaked at saddles rather than minima
by extending the range of $r$ to label all stationary points and defining the corresponding
states $\psi^{(r)}_0$ to be
\be 
\psi^{(r)}_0=p^{(r)}_0(x)|0\ra~~,
\ee
where
\be
p^{(r)}_0(x)
=\fr{1}{\sqrt{Z^{(r)}}}\exp\left\{-\fr{\bb}{2}[B_r
                   +\fr{1}{2}(x_m-\xb_m^{(r)})A^{(r)}_{mn}(x_n-\xb^{(r)}_n)]\right\}~~,
\ee
and
\be
A^{(r)}_{mn}=\sum_\gg |\ll^{(r)}_\gg|c^{(r)}_{\gg m}c^{(r)}_{\gg n}~~,
\label{QUAD}
\ee
$\ll^{(r)}_\gg$ being the eigenvalues (not all positive) of $\phi_{mn}(\xb^{(r)})$
and $c^{(r)}_{\gg n}$ the corresponding eigenvectors. This form for these eigenstates 
implies that it is not very convenient to express them in the "typical" form of eq(\ref{typical}).
Of course the higher excitations based on the quadratic structure associated with $A^{(r)}_{mn}$
in eq(\ref{QUAD}) are also available as eigenstates.

\subsection{\label{Fermi}Fermionic Levels}
 
An analysis of low lying states at the first fermionic level can be carried out along the same lines.
As pointed out by Witten \cite{WITTEN} and further discussed by 
T{\v a}nase-Nicola and Kurchan \cite{KURCHAN1}, the relevant wavefunctions are those concentrated
at the stationary points $\xb^{(r)}$ of $\phi(x)$ for which $\phi_{mn}(\xb^{(r)})$ has one negative
eigenvalue, that is at saddle points with a Morse index of unity. Indeed they contain as factors
the saddle wavefunctions discussed in subsection \ref{Zero}. The term $\phi_{mn}(\xb^{(r)})a^\dag_ma_n$ in the 
supersymmetric Hamiltonian appropriate to this saddle, guarantees that the lowest lying state 
has zero eigenvalue in this approximation. Wavefunctions associated with excitations at
these saddles have eigenvalues proportional to $\bb$, The other wavefunctions based on the original 
minima and on saddles with a Morse index of two or higher are also available as eigenfunctions 
but have lowest eigenvalues proportional to $\bb$~. We will not repeat the analysis in detail 
but mention two points. 

First, since the $R-1$ low lying eigenstates in $\Hc_0^{(+)}$ 
can be raised by the application of $Q^\dag$ to provide the low lying eigenstates
in $\Hc_1^{(-)}$ there must be at least $R-1$ saddle points of $\phi(x)$ with unit Morse index.
If there are more than $R-1$ such saddles then the excess is accounted for by the
corresponding number of low lying eigenstates in $\Hc_1^{(+)}$. By again applying
$Q^\dag$ to these states we create a number of low lying eigenstates in $\Hc_2^{(-)}$
which then in turn provide a lower bound on the number of saddles of $\phi(x)$
with a Morse index of 2. The argument can be carried on until we run out of higher
saddles or maxima. This argument is the basis on which one can derive
the Morse inequalities from supersymmetry \cite{WITTEN,KURCHAN1}.

Second the action of $Q^\dag$ on the approximate form for the low lying
eigenstates of, say, $\Hc_0^{(+)}$ produces zero because the eigenvalue of
each of these states is zero in this approximation. It follows that the 
approximation is inadequate for dealing with the subtle structure of the
eigenstates that allows $Q^\dag$ to move the peaks in the wavefunctions
in $\Hc_0^{(+)}$ at the minima of $\phi(x)$ to the peaks of the 
wavefunctions in $\Hc_1^{(-)}$ at the saddles of $\phi(x)$~. 

\subsection{Current Flow at Large Times}

As emphasised by T{\v a}nase-Nicola and Kurchan \cite{KURCHAN1,KURCHAN2}
the way to understand the link between fermionic levels is to study the currents associated
with the states in $\Hc_0^{(+)}$~. Any initial probability distribution can be expressed
as a superposition of eigenfunctions of $\hat{H}_0$ (see eq(\ref{ham1})). 
After a sufficient length of time the influence of contributions from
eigenfunctions associated with high lying eigenvalues will have died away
and the system will acquire a probability distribution that lies in the 
subspace spanned by the set $P_0^{(r)}(x)$~. As time progresses the occupation
of the various minima will settle down to the equilibrium values. The equilibriating
process is the flow between minima of the currents mentioned above \cite{KURCHAN1,KURCHAN2}.
For our purposes then the system starts in a state $\psi$ where
\be
\psi=V(x)p_0(x)\simeq \sum_r V_rw_rp_0^{(r)}(x)~~,
\ee
where $V_r=V(\xb^{(r)})$~.
This corresponds to the probability distribution
\be
P(x)=V(x)P_0(x)\simeq \sum_r V_rw_r^2P_0^{(r)}(x)~~.
\ee
We assume that $V(x)$ is, in the neighbourhoods of each of the minima, 
a relatively smooth function after the initial fast
partial relaxation. Now by exploiting our electromagnetic analogy we can identify a 
charge, $q_r$, associated with the minimum at $\xb^{(r)}$,
\be
q_r=\int dxV_rw_r^2P_0^{(r)}(x)=w_r^2V_r~~,
\ee
and we can identify $V_r$ as the voltage at $\xb^{(r)}$, then we see that $w_r^2$
is the capacitance at that minimum. Now
\be
\fr{\d q_r}{\d t}=-\sum_s I_{rs}~~,
\ee
where $I_{rs}$ is the total current flowing from $\xb^{(r)}$ to $\xb^{(s)}$~.
Because of the linearity of the diffusion process we have
\be
I_{rs}=R^{-1}_{rs}(V_r-V_s)~~,
\ee
where $R_{rs}$ is the resistance between the two minima. As a result the time dependence
of the equations for the voltages becomes
\be
w_r^2\fr{\d V_r}{\d t}=-\sum_s R^{-1}_{rs}(V_r-V_s)~~.
\label{volts}
\ee
We can then find the low lying eigenvalues of $H_0$ by looking for the 
solutions of this equation that have the form $V_r=U_re^{-pt}$ where
$U_r$ are constants that obey the homogeneous equations
\be
w_r^2pU_r=\sum R^{-1}_{rs}(U_r-U_s)~~.
\label{ELEC}
\ee
Of course $p=TE$, where the values of $E$ that make these equations soluble,
are the low-lying eigenvalues of $H_0$. 

The problem then reduces to calculating the resistances.
Recall that the principle of minimum energy dissipation 
tells us that the actual current distribution is along paths of least resistance.
The local resistivity is $\bb/P_0(x)$ and is least in the neighbourhoods of
paths of steepest descent (where these exist) passing between minima and over
saddles (Morse index 1). When $\bb$ is very large any departure from these 
neighbourhoods penalises the flow of current in terms of energy dissipation 
very severely.

Consider a typical pair of minima connected across a saddle at $x_S$~.
Eq(\ref{newdiff1}) tells us that away from the sources of current at the minima
the current distribution is divergenceless. This confirms that the total current
flowing parallel to the curve of steepest descent is constant along the curve.
Near a point $x_L$ on the curve of steepest descent with unit tangent vector $\bt$ 
the current distribution is $\bj(x)\propto \bt$~. By definition 
\be
\bj=-TP_0(x)\nabla V(x)~~.
\ee
This tells us that $\nabla V(x)\propto \bt$ near the curve of steepest descent
and hence that $V(x)$ is constant across the profile of the current. The 
total current is $I$ where
\be
I=\int dx_\perp \bj\cdot\bt~~,
\ee
$x_\perp$ are a set of two Cartesian coordinates orthogonal to the curve at $x_L$
and $dx_\perp$ is the transverse area element.

Now
\be
P_0(x)=P_0(x_L)\exp\left\{-\fr{1}{2}\bb x_\perp^TA_\perp(x_L)x_\perp\right\}~~,
\ee
for points near $x_L$~. Here $A_\perp(x_L)$ is the $2\times2$ matrix
determining the locally quadratic cross-section of the "wire" conducting the
current in the neighbourhood of the curve of steepest descent from the saddle. 
We can define a cross-section value for the wire, $\Ac(x_L)$ through the 
equation
\be
\int dx_\perp P_0(x)\simeq P_0(x_L)\Ac(x_L)~~,
\ee
That is
\be
\Ac(x_L) =\left(\fr{2\pi}{\bb}\right)\fr{1}{\sqrt{\det A_\perp(x_L)}}=\left(\fr{2\pi}{\bb}\right)\fr{1}{\sqrt{\gg_1\gg_2}}~~,
\ee
where $\gg_1$ and $\gg_2$ are the positive eigenvalues of the matrix $A(x_L)=\{\phi_{mn}(x_L)\}$ and hence also
of $A_\perp(x_L)$~.
It follows then that
\be
I=-TP_0(x_L)\Ac(x_L)\bt\cdot\nabla V(x_L)~~,
\ee
or
\be
\bt\cdot\nabla V(x_L)=\fr{\d}{\d s}V(x_L)=-\bb\fr{1}{P_0(x_L)\Ac(x_L)}I~~,
\label{res1}
\ee
where $s$ is distance along the curve of steepest descent.
On integrating along the curve of steepest descent between the minima
we see that the resistance to the flow of current is
\be
R=\bb\int ds \fr{1}{P_0(x_L)\Ac(x_L)}~~,
\label{res2}
\ee
where the integration covers the path between the minima.
However in the limit of large $\bb$ the integral is dominated by the contribution
from near the saddle at the point $x_S$ where the integrand is largest. We have then in this limit 
\be
R=\fr{\bb}{P_0(x_S)\Ac(x_S)}\int ds e^{-\fr{1}{2}\bb\gg_{||} s^2}
                   =\fr{\bb}{P_0(x_S)\Ac(x_S)}\sqrt{\fr{2\pi}{\bb\gg_{||}}}~~,
\label{resist}
\ee
where $-\gg_{||}$ is the negative eigenvalue of the matrix $A(x_S)$ with components $\phi_{mn}(x_S)$~.
In the same way we can calculate the resistance between any two minima of $\phi(x)$~.

Later we will require the above result specialised to one and two dimensions. The one-dimensional result 
is obtained by setting $\Ac(x_L)=1$ and we have
\be
R=\fr{1}{P_0(x_S)}\sqrt{\fr{2\pi\bb}{\gg_{||}}}~~.
\label{resist1d}
\ee
The two-dimensional result is obtained by setting the one-dimensional cross-section 
$\Ac(x_L)$ to
\be
\Ac(x_L)=\sqrt{\fr{2\pi}{\bb\gg_1}}~~,
\ee
where $\gg_1$ is the single positive eigenvalue of $A(x_S)$~. We have
\be
R=\fr{\bb}{P_0(x_S)}\sqrt{\fr{\gg_1}{\gg_{||}}}
\label{resist2d}
\ee

\subsection{\label{onefer}One-Fermion Level}

The behaviour of the system at the one-fermion level parallels that at the 
zero-fermion level. As discussed in subsection(\ref{OF1}) the wave functions 
in $\Hc_1^{(+)}$ are characterised by a field distribution $\Hb(x)$ and an associated field 
$\Bb(x)=P_0(x)\Hb(x)$ satisfying $\nabla.\Bb(x)=0$~.
That part of an initial field distribution represented by eigenfunctions with high lying
eigenvalues will decay rapidly as will the corresponding contributions to the 
energy $U=1/2\int dx\Bb^2/P_0(x)$. 
The remaining contributions from the eigenstates with low lying eigenvalues
will survive for much longer and will be associated with field distributions $\Bb$ that
permit the energy $U$ to be as low as possible in the circumstances. In the  
zero-fermion sector this argument implied that the density retreated to the neighbourhoods of the 
minima of $\phi(x)$~. In the one-fermion sector, because the field $\Bb$ has
zero divergence, it cannot be concentrated only in the disjoint neighbourhoods of the minima of
$\phi(x)$ but must instead be distributed along the filaments surrounding the
paths of steepest descent joining these minima across saddles with
Morse index one. In other words $\Bb$ lies in the same channels as the currents
we discussed in the zero-fermion sector. Since $\Bb$ is a conserved field the 
actual distribution must be a superposition of loops of field. This is the same picture as 
elucidated in \cite{KURCHAN1,KURCHAN2} except that in our
electromagnetic analogy we replace the conserved current with the magnetic induction field.
The number of independent loops is of course equal to the number of saddles with Morse index two.

For simplicity consider a situation with only one saddle of Morse index two and one
associated loop $L$. We define a magnetic loop integral, $\Mc$, by 
\be
\Mc=\int_L d\bx\cdot\Hb~~,
\ee
where the integration is round the loop $L$. As argued above, in the limit of large $\bb$, and
for times sufficiently large, the $\Bb$ field distribution will closely follow the 
loop of curves of steepest descent as will therefore the field $\Hb$ in the neighbourhood
of the loop. If we integrate, as in the resistance calculation in the above discussion of the 
zero-fermion sector, over the 
cross-section of the loop current then we find that the total magnetic flux, $B$, which is constant round 
the loop, is related to the magnetic field at the loop by
\be
B\bt=P_0(x_L)\Ac(x_L)\Hb(x_L)~~,
\ee
where $\bt$ is the unit tangent vector at $x_L$ on the loop.
That is, along the loop,
\be
\Hb(x_L)=\fr{1}{P_0(x_L)\Ac(x_L)}B\bt~~,
\ee
and therefore on integrating round the loop we have
\be
\Mc=TR_LB~~,
\label{EMFLOOP1}
\ee
where $R_L$ is identical to the total resistance round the loop evaluated as the sum of the resistances
of the various segments evaluated previously. 

To complete the dynamics we apply eq(\ref{integ2}) by choosing the surface $S$
to be a disc-like element crossing the channel of steepest descent in which the 
magnetic flux field is concentrated. This implies that
\be
\fr{\d B}{\d t}=-\int_C\Eb\cdot d\bx~~.
\label{integ4}
\ee
Note that this is true for {\it any} loop $C$ encircling the the steepest descent curve.
The result is therefore independent of the position $\bx_{SD}$ at which the loop $C$ cuts the 
surface of steepest descent emerging from the saddle with Morse index 2. It is also independent
of the angle at which it crosses the surface. That this is possible is due to the fact that 
$\Eb=TP_0(x)\bj$ is significant only in the neighbourhood of the surface of steepest descent, $\bj$
being smooth in this neighbourhood. Independence of the angle of incidence requires that 
$\Eb$ and $\bj=\nabla\times\Hb$ are orthogonal to the surface of steepest descent. 
On evaluating the right side of eq(\ref{integ4}), in the limit of high $\bb$, we find
\be
\fr{\d B}{\d t}=-TP_0(x_{SD})\Tc(x_{SD})\bn.\nabla\times\Hb(x_{SD})~~,
\label{integ5}
\ee
where $\Tc(x_{SD})$ is the thickness of the steepest descent surface defined so that
\be
\int \bn.d\bx P_0(x)=P_0(x_{SD})\Tc(x_{SD})~~,
\ee
where $\bn$ is the normal to the surface of steepest descent at $x_{SD}$.
If we make explicit the quadratic structure of the transverse dependence of $\phi(x)$
then
\be
P_0(x)=P_0(x_{SD})e^{-\fr{1}{2}\bb\aa_\perp(\bn.\bx)^2}~~,
\ee
and
\be
\Tc(x_{SD})=\sqrt{\fr{2\pi}{\bb\aa_\perp}}~~,
\ee
where $\aa_\perp$ is the second derivative of $\phi(x)$ in the direction of $\bn$.

On the surface of steepest descent we have from eq(\ref{integ4})
\be
\bn.\nabla\times\Hb(x_{SD})=-\fr{\bb}{P_0(x_{SD})\Tc(x_{SD})}\fr{\d B}{\d t}~~.
\ee
Integrating both sides of this equation over the surface of steepest descent
we obtain
\be
\Mc=-\int_{SD}dS\fr{\bb}{P_0(x_{SD})\Tc(x_{SD})}\fr{\d B}{\d t}~~.
\label{integ6}
\ee
In the limit of large $\bb$ the integral on the right is dominated by
$x_{S_2}$, the saddle with Morse index 2. We have 
\be
\Mc=-\sqrt{\fr{2\pi}{\bb\aa_1}}\sqrt{\fr{2\pi}{\bb\aa_2}}\fr{\bb}{P_0(x_{S_2})\Tc(x_{S_2})}\fr{\d B}{\d t}~~.
\label{integ7}
\ee
Here $\aa_1$ and $\aa_2$ are the eigenvalues of the matrix of second derivatives of $\phi(x)$
along the surface at the saddle point.

Combining eq(\ref{integ7}) and eq(\ref{EMFLOOP1}) we find
\be
\fr{\d B}{\d t}=-\fr{1}{2\pi}TR_LP_0(x_{S_2})\Tc(x_{S_2})\sqrt{\aa_1\aa_2}B~~.
\ee
From this equation we can read off the decay exponent $\nu$. It is
\be
\nu=\fr{1}{2\pi}TR_LP_0(x_{S_2})\Tc(x_{S_2})\sqrt{\aa_1\aa_2}=TR_LP_0(x_{S_2})\sqrt{\fr{\aa_1\aa_2}{2\pi\bb\aa_\perp}}~~.
\label{floop}
\ee
This result can be generalised to deal with situations involving more than
one saddle with Morse index two and hence more than one loop. 

The result can be specialised to the two-dimensional case by setting the 
thickness $\Tc(x_{SD})$ to unity with the result
\be
\nu=\fr{1}{2\pi}TR_LP_0(x_{S_2})\sqrt{\aa_1\aa_2}~~.
\label{floop2d}
\ee
\section{\label{einstein}External Electric Field}

We can apply our understanding of the supersymmetry structure to
the case of an external electric field and the calculation of
the associated susceptibilities. In particular we derive the 
appropriate form of the Einstein relation between the susceptibility
matrix and the fluctuation correlation matrix. 

From subsection {\ref{extfld}} we see that we are dealing with 
a field of the form $f_{na}(X)=\d_nf_a(X)$~.
It follows that 
\be
Q^\dag f_{ma}(x)a^\dag_m\psi_0=0~~.
\ee
From eq(\ref{diff8a}) and the fact that $Q^\dag$ commutes with the Hamiltonian $H$
we find
\be
HQ^\dag\psi_a=0~~.
\ee
Now $Q^\dag\psi_a\in\Hc_2$ and $H$ has no zero eigenvalues in this subspace
it follows that 
\be
Q^\dag\psi_a=0~~,
\label{ann3}
\ee
and therefore that $\psi_a\in \Hc_1^{(-)}$~. 

We have also
\be
f_{ma}(x)a^\dag_m\psi_0=iQ^\dag f_a(x)\psi_0~~.
\ee
Combining this result with eq(\ref{diff8a}) and eq(\ref{ann3}) we find that 
\be
Q^\dag(Q\psi_a-i\bb f_a(x)\psi_0)=0~~.
\ee
It follows that
\be
Q\psi_a-i\bb f_a(x)\psi_0=-i\bb c_a\psi_0~~,
\ee
for some coefficients $\{c_a\}$~.
Taking a scalar product with $\psi_0$ we obtain the result
\be
c_a=(\psi_0,f_a(x)\psi_0)=\int dxP_0(x)f_a(x)={\bar{f}}_a~~.
\ee
We have then 
\be
Q\psi_a=i\bb (f_a(x)-\fb_a)\psi_0~~.
\label{susy7}
\ee
Multiplying by $(f_b(x)-\fb_b)$ and taking the scalar product with $\psi_0$  
we find
\be
((f_b(x)-\fb_b)\psi_0,Q\psi_a)=i\bb\SS_{ab}~~,
\label{ER1}
\ee
where
\be
\SS_{ab}=\la(f_a(x)-\fb_a)(f_b(x)-\fb_b)\ra~~.
\ee
The left side of eq(\ref{ER1}) can be put in the form
\be
(Q^\dag(f_b(x)-\fb_b)\psi_0,\psi_a)=(-i\d_nf_a(x)a^\dag_n\psi_0,F_{ma}(x)a_m^\dag\psi_0)~~.
\ee
The right side of this equation can be expressed as
\be
i\int dxP_0(x)(\d_mf_a(x))F_{ma}(x)=i\la \xi_{ma}\d_mf_b(x)\ra~~.
\label{ER2}
\ee
Combining eq(\ref{ER1}) and eq(\ref{ER2}) we obtain
\be
\la \xi_{ma}\d_mf_b(x)\ra=\bb\SS_{ab}~~.
\ee
This is our general form for the Einstein relation. In this form it can be generalised to
arbitrary, possibly compact, manifolds.

If we specialise to the case for which $f_a(x)=x_a$ we obtain the 
more standard result, relevant to a {\it uniform} external field,
\be
\xib_{ab}=\bb\la(x_a-\xb_a)(x_b-\xb_b)\ra~~.
\ee
The argument given here is a generalisation of the corresponding argument
for a one-dimensional system presented in \cite{DEAN}.

\subsection{\label{compsus}Computation of Susceptibility}

We are particularly interested in the susceptibility of the system in a uniform external field.
Note that there is no distinction here between the labels $\{a,b,...\}$ and $\{m,n,...\}$~.
To compute this susceptibility we evaluate the conditional susceptibility $F_{mn}(x)$
in terms of the eigenstates of $H$ that lie in $\Hc_0^{(+)}$~.

Eq(\ref{ann3}) implies that 
\be
\psi_n=iQ^\dag\chi_n~~,
\ee
for some $\chi_n\in\Hc_0^{(+)}$~. If we set $\chi_n=V_n(x)\psi_0$ then
\be
F_{mn}(x)=\d_mV_n(x)~~.
\label{locsus1}
\ee
Although $F_{mn}(x)$ is unchanged by the addition of a constant to $V_n(x)$, 
in fact $V_n(x)$ and hence $\chi_n$ are rendered unique by the requirement that
$\chi_n$ is orthogonal to $\psi_0$~. Eq(\ref{susy7}) implies that
\be
QQ^\dag\chi_n=\bb(x_n-\xb_n)\psi_0~~.
\ee
That is
\be
H\chi_n=\bb(x_n-\xb_n)\psi_0~~
\label{FMN1}
\ee
Again denote the eigenstates of $H$ in $\Hc_0^{(+)}$ with eigenvalues $E_\aa$, 
by $v_\aa(x)\psi_0$. We have then
\be
\chi_n=\sum_\aa C_{\aa n}v_\aa(x)\psi_0~~.
\ee
where 
\be
C_{\aa n}=(v_\aa\psi_0,\chi_n)~~.
\ee
From eq(\ref{FMN1}) we have
\be
C_{\aa n}=\fr{1}{E_\aa}(v_\aa\psi_0,\bb(x_n-\xb_n)\psi_0)
          =\fr{\bb}{E_\aa}d_{\aa n}~~,
\ee
where $d_{\aa n}$ is the dipole moment associated with the state $v_\aa(x)\psi_0$
and is given by
\be
d_{\aa n}=\int dxP_0(x)v_\aa(x)(x_n-\xb_n)~~.
\label{dipole}
\ee
We have then
\be
V_n(x)=\sum_\aa v_\aa(x)\fr{\bb}{E_\aa}d_{\aa n}~~.
\label{locsus2}
\ee
From $V_n(x)$ we can compute $F_{mn}(x)$~. 
We have
\be
P_0(x)F_{mn}(x)
 =\sum_\aa j^{(\aa)}_m(x)\fr{\bb}{E_\aa}d_{\aa n},
\label{sus1}
\ee
where $j^{(\aa)}_m(x)$ is a current associated with $v_\aa(x)$ and is given by
\be
j^{(\aa)}_m=P_0(x)\d_mv_\aa(x)~~.
\label{sus2}
\ee
The eigenvalue equation for $v_\aa(x)$ is equivalent to the result
\be
\d_mj^{(\aa)}_m(x)=-E_\aa v_\aa(x)P_0(x)~~.
\ee
This can be re-expressed in the form
\be
j^{(\aa)}_m(x)=E_\aa(x_m-\xb_m)v_\aa(x)P_0(x)+\d_k[(x_m-\xb_m)j^{(\aa)}_k(x)]~~,
\ee
where the introduction of the term with $\xb_m$ is arbitrary but convenient.
We have then
\be
P_0(x)F_{mn}(x)=\bb\sum_\aa(x_m-\xb_m)v_\aa(x)P_0(x)d_{\aa n}+\d_kY_{kmn}~~.
\label{sus3}
\ee
where
\be
Y_{kmn}= \sum_\aa\fr{\bb}{E_\aa}(x_m-\xb_m)j^{(\aa)}_k(x)d_{\aa n}~~.
\ee
On integrating both sides of eq(\ref{sus3}) the total divergence drops out and we obtain
\be
\xib_{mn}=\bb\sum_\aa d_{\aa m}d_{\aa n}~~.
\label{sus4}
\ee
The result is, as it should be, symmetrical in $m$ and $n$ even though
this symmetry is not required for $F_{mn}(x)$ itself. Note that the
summation over $\aa$ can be extended to include the null contribution
from the state $\psi_0$~.

\subsection{\label{SUSC}Derivation of Susceptibility at High $\bb$}

In eq(\ref{sus4}) the summation $\sum_\aa$ can be split into two parts.
The first part which we denote also by $\sum_\aa$, has $\aa$ restricted to
the low lying eigenstates of $H_0$ including the ground state $\psi_0$~.
If we denote this contribution by $\xib^{(1)}_{mn}$ then eq(\ref{sus4}) tells us
that
\be
\xib^{(1)}_{mn}\simeq\bb\sum_{\aa rr'}w_r^2v_\aa(\xb^{(r)}_m)(\xb^{(r)}_m-\xb_m)
                             w_{r'}^2v_\aa(\xb^{(r')})(\xb^{(r')}_n-\xb_n)~~.
\ee
If we exploit the orthogonality of the matrix with elements $\{w_r^2v_\aa(\xb^{(r)})\}$ 
we have 
\be
\sum_\aa w_r^2v_\aa(\xb^{(r)}_m)w_{r'}^2v_\aa(\xb^{(r')})=w_r^2\dd_{rr'}~~,
\ee
and ultimately
\be
\xib^{(1)}_{mn}\simeq\bb\sum_r w_r^2(\xb^{(r)}_m-\xb_m)(\xb^{(r)}_n-\xb_n)~~.
\ee

The second part of the summation is over high lying eigenstates of $H_0$~.
We can describe this summation by making the replacement $\aa\rightarrow (r,\gg)$
where $r$ runs over the set of minima and $\gg$ runs over the high lying
states of the quadratic approximation to the $r$-th minimum. In order to
compute the local dipole moment at the $r^{\hbox{th}}$ minimum we can restrict 
the summation over $\gg$ taking into account only those states 
account for the lowest set of such states for which the corresponding $v_\aa(x)$
is linear in $x$~. It is then reasonable to
assume that the mixing between the neighbourhoods of different
minima experienced by yet higher lying states can be ignored. 

There is a subtlety relating to the normalisation of the states. We have
assumed that
\be
\int dxP_0(x)v_\aa(x)v_{\aa'}(x)=\dd_{\aa\aa'}~~.
\ee
For states concentrated in the neighbourhood of the $r$-th minimum
this means that
\be
\int dxw_r^2P^{(r)}_0(x)v_\aa(x)v_{\aa'}(x)=\dd_{\aa\aa'}~~.
\ee
If instead we denote these localised states by $v^{(r)}_\gg(x)$ with the normalisation
\be
\int dx P^{(r)}_0(x)v^{(r)}_\gg(x)v^{(r)}_{\gg'}(x)=\dd_{\gg\gg'}~~,
\ee 
then we must reconcile the two normalisations by making the replacement
$v_\aa(x)\rightarrow w_r^{-1}v^{(r)}_\gg(x)$ in eq(\ref{dipole})~.
Denoting then the second contribution to the right side of eq(\ref{sus4}) we have
\be
\xi^{(2)}_{mn}\simeq\bb\sum_{r\gg}w_r^2\int dxP^{(r)}_0(x)v^{(r)}_\gg(x)(x_m-\xb_m)
                    \int dx'P^{(r')}_0(x')v^{(r)}_\gg(x')(x'_n-\xb_n)~~.
\ee
Finally we can make the replacement $\xb_n\rightarrow\xb^{(r)}_n$ in this equation 
without changing the result. It follows then that
\be
\xi^{(2)}_{mn}\simeq\bb\sum_rw_r^2\SS^{(r)}_{mn}~~.
\ee
The result is
\be
\xi_{mn}=\bb\sum_rw_r^2\left\{(\xb^{(r)}_m-\xb_m)(\xb^{(r)}_n-\xb_n)+\SS^{(r)}_{mn}\right\}~~,
\ee
and is consistent with the Einstein relation if we identify the variance matrix as
$\SS_{mn}$ where
\be
\SS_{mn}=\sum_rw_r^2\left\{(\xb^{(r)}_m-\xb_m)(\xb^{(r)}_n-\xb_n)+\SS^{(r)}_{mn}\right\}~~.
\label{variance}
\ee
It should be noted however that the terms $\SS^{(r)}_{mn}$ are $O(1/\bb)$. In order 
for the asymptotic expansion to be consistent we should include all effects
of this order. These are accounted for by reinterpreting $\xb^{(r)}_m$ as the mean position
of a particle that is in the neighbourhood of the $r^{\mbox{th}}$ minimum rather than
the position of the minimum itself. The two positions differ by terms that are $O(1/\bb)$~.
This difference comes about from an approximation to the ground state wave function
that goes beyond the quadratic approximation to $\phi(x)$ and includes {\it third} order 
derivatives of $\phi(x)$ in its Taylor expansion. We will encounter examples of this 
in section \ref{onedex} and section \ref{twodex}~.
With this adjustment the result is indeed intuitively plausible and in this form it can be 
derived directly from the partition function in the high-$\bb$ limit. 

\subsection{\label{Areasus}Area Susceptibility}

Evaluating a higher order susceptibility such as the area susceptibility
is a more complicated task which we do not pursue here in detail. However,
it is of interest to investigate the possibility of a generalisation of the
Einstein relation that holds for the linear susceptibility. We follow a line of 
reasoning for the area susceptibility
that is as close as possible to that for the original Einstein relation.
The result however is not as useful for reasons we explain.
The starting point for the analysis is eq(\ref{diff9a})~.
We continue to work with the electric field case so we have from the right side of eq(\ref{diff9a}),
\begin{eqnarray}
f_{ma}(x)\psi_b-f_{mb}(x)\psi_a&=&((\d_mf_a(x))\psi_b-(\d_mf_b(x))\psi_a)\\\nonumber
                           &=&iQ^\dag((f_a(x)-\fb_a)\psi_b-(f_b(x)-\fb_b)\psi_a)~~,
\end{eqnarray}
where we have used the fact that $Q^\dag\psi_a=0$. It then follows that
\be
Q^\dag(f_{ma}(x)\psi_b-f_{mb}(x)\psi_a)=0~~.
\ee
Since $Q^\dag$ commutes with the Hamiltonian $H$ we see from eq(\ref{diff9a}) that 
\be
HQ^\dag \psi_{ab}=0~~,
\ee
and hence that
\be
Q^\dag \psi_{ab}=0~~.
\ee
Eq(\ref{diff9a}) can now be expressed in the form
\be
Q^\dag(Q\psi_{ab}-i\bb((f_a(x)-\fb_a)\psi_b-(f_b(x)-\fb_b)\psi_a))=0~~.
\ee
That is
\be
Q\psi_{ab}-i\bb((f_a(x)-\fb_a)\psi_b-(f_b(x)-\fb_b)\psi_a)=Q^\dag\pi_{ab}~~,
\label{asus1}
\ee
where $\pi_{ab}=g_{ab}(x)\psi_0\in\Hc_0^{(+)}$~. We have
\be
H\pi_{ab}=QQ^\dag\pi_{ab}=-i\bb Q((f_a(x)-\fb_a)\psi_b-(f_b(x)-\fb_b)\psi_a)~~.
\label{asus2}
\ee
Eq(\ref{asus2}) is sufficient to determine $\pi_{ab}$ in terms of $f_a(x)$ and
$\psi_a$~. 
We take the scalar product of each side of eq(\ref{asus1}) with
$$
(f_c(x)-\fb_c)Q^\dag(f_d(x)-\fb_d)\psi_0~~.
$$
It is simple to see that the left side can be put in the form
\be
(Q^\dag(f_c(x)-\fb_c)Q^\dag(f_d(x)-\fb_d)\psi_0,\psi_{ab})
             =-\la(\xi_{ma}\xi_{nb}-\xi_{mb}\xi_{na})f_{mc}(x)f_{nd}(x)\ra=-\chi_{ab,cd}~~,
\ee
where $\chi_{ab,cd}$ is the area susceptibility tensor. However when we complete the
manipulation by forming the appropriate scalar product with the right side of eq(\ref{diff9a})
we obtain after some tedious algebra two contributions. 
The first is
$$
\fr{1}{2}\bb\la(f_{kc}(x)(f_d(x)-\fb_d)-f_{kd}(x)(f_c(x)-\fb_c))
                   (\xi_{kb}(f_a(x)-\fb_a)-\xi_{ka}(f_b(x)-\fb_b))\ra~~,
$$
and the second is
$$
-\fr{1}{2}\la(\d_kg_{ab}(x))((f_{kc}(x))(f_a(x)-\fb_a)-(f_{kd}(x))(f_c(x)-\fb_c))\ra~~.
$$
These expressions are not particularly helpful. We have presented them however in order
to underline the point that in this case there is no expression for the area susceptibility
that parallels the Einstein relation for the linear susceptibility and involves only the distribution $P_0(x)$~. 
Instead the two terms above require a knowledge of the
complete joint equilibrium distribution $P(x,\xi)$~. This reduces the usefulness of the
relation and means that the only way to approach the numerical evaluation of the area susceptibility 
is through the full simulation of the slave equations. We show some results later for
a specific case in section \ref{twodex}~.

\section{\label{onedex}One-Dimensional Examples}

There are also some general issues that can be addressed particularly easily in one dimension.
In a previous paper \cite{DEAN} we examined the the case of a double minimum profile
and emphasised the power-law probability distribution acquired by the slave variable
$\xi$ at large values. The origin of this phenomenon lies in the existence of regions
on $x$-space for which $\phi''(x)<0$ \cite{DEAN,POPE,DRU3,SCH}. 

Such a power-law distribution raises the possibility that even though the mean of
$\xi$ exists its variance may diverge together with moments higher than the second.
Since we are interested in comparing numerical results with asymptotic calculations 
at small $T$ this phenomenon, which becomes increasingly severe in this limit, affects
how we assess the outcome of our simulation calculation.  We therefore develop a criterion for 
predicting the appearance of a diverging variance, in the one dimensional context.

In the limit of large time the resulting static distribution $P(x,\xi)$  satisfies
\be
\fr{\d}{\d x}\left(\fr{\d}{\d x}+\bb\phi'(x)\right)P(x,\xi)
                                   +\bb\fr{\d}{\d \xi}\left(\phi''(x)\xi-1\right)P(x,\xi)=0~~.
\label{critp1}
\ee
The support for $P(x,\xi)$ lies in $\xi >0$ with the boundary condition $P(x,0)=0$~.
If we define $Q(x,p)$ as
\be
Q(x,p)=\int_0^{\infty}d\xi~\xi^p P(x,\xi)~~.
\ee
then we know that $Q(x,p)$ is convergent for $0\le p <p_c$ where for $p>p_c$ $Q(x,p)$
is divergent. Since $Q(x,1)$ is convergent we know that $p_c>1$~. 

From eq(\ref{critp1}) we find
\be
\fr{\d}{\d x}\left(\fr{\d}{\d x}+\bb\phi'(x)\right)Q(x,p)
                                     -p\bb\phi''(x)Q(x,p)+p\bb Q(x,p-1)=0~~.
\label{critp2}
\ee
If we set $Q(x,p)=q(x,p)p_0(x)$ then eq(\ref{critp2}) becomes
\be
H_pq(x,p)=p\bb q(x,p-1)~~,
\ee
where
\be
H_p=-\fr{\d^2 }{\d x^2}+\fr{1}{4}\bb^2(\phi'(x))^2+(p-\fr{1}{2})\bb\phi''(x)~~.
\label{critp3}
\ee
For $1\le p$ we can solve for $q(x,p)$ in terms of $q(x,p-1)$ provided $H_p$
does not have a zero eigenvalue. This will remain true for $p<p_c$. The critical value
$p_c$ is determined by the condition that $H_p$ has a zero eigenvalue when $p=p_c$~.
There are straightforward numerical techniques such as Sturm sequencing for computing
the eigenvalues of $H_p$. By varying $p$ through an appropriate range we can determine the
value of $p$ that yields a zero eigenvalue. If this critical value is less than two then
we are in a situation where the variance of the distribution for $\xi$ is divergent.

We can see directly how a power law distribution for $\xi$ induces a critical value for $p$.
Define the probability distribution function $D(\xi)$ so that
\be
D(\xi)=\int dxP(x,\xi)~~.
\ee
Our claim is that for large $\xi$
\be
D(\xi)\simeq \fr{A}{\xi^{\aa+1}}~~,
\label{expnt}
\ee
for some constant $A$ and exponent $\aa$.
If we set
\be
D(\xi,p)=\int_0^\xi d\xi'\xi'^pD(\xi')~~,
\ee
then, assuming the above asymptotic behaviour for $D(\xi)$ we see
that
\be
D(\xi,p)\simeq \bar{\xi^p}-\fr{A}{\aa-p}\fr{1}{\xi^{\aa-p}}~~.
\label{Dasymp}
\ee
It follows that the critical value at which $\bar{\xi^p}$ diverges is $p_c=\aa$~. We
can use this result to compute $\aa$ and compare with the results of our simulation.

\subsection{\label{mex1}Double Minimum Profile}

The Double Minimum Profile occurs in a model for which
\be
\phi(x)=\fr{\ll}{4}(x^2-a^2)^2~~.
\label{dminprof}
\ee
Clearly $\phi(x)$ has minima at the points $x=a$ and $x=-a$~. We label them $r=1$ and $r=2$
respectively. A maximum of $\phi(x)$ lies at the origin. We note that $\phi(\pm a)=0$,
$\phi(0)=\ll a^4/4$ and $\phi''(\pm a)=2\ll a^2$, $\phi''(0)=-\ll a^2$~. In fact $\phi''(x)<0$
for $-a/\sqrt(3)<x<a/\sqrt(3)$~. It follows that we expect the slave variable to exhibit a power-law
probability distribution at large values of $\xi$~. 

\subsubsection{\label{dmcritp1}Critical Moment for Double Minimum}

For this model the Hamiltonian $H_p$ in eq(\ref{critp3}) becomes
\be
H_p=-\fr{\d^2}{\d x^2}+\fr{1}{4}\bb^2\ll^2 x^2(x^2-a^2)^2+(p-\fr{1}{2})\bb\ll(3x^2-a^2)~~.
\ee
\begin{figure}[t]
   \centering
  \includegraphics[width=0.8\linewidth]{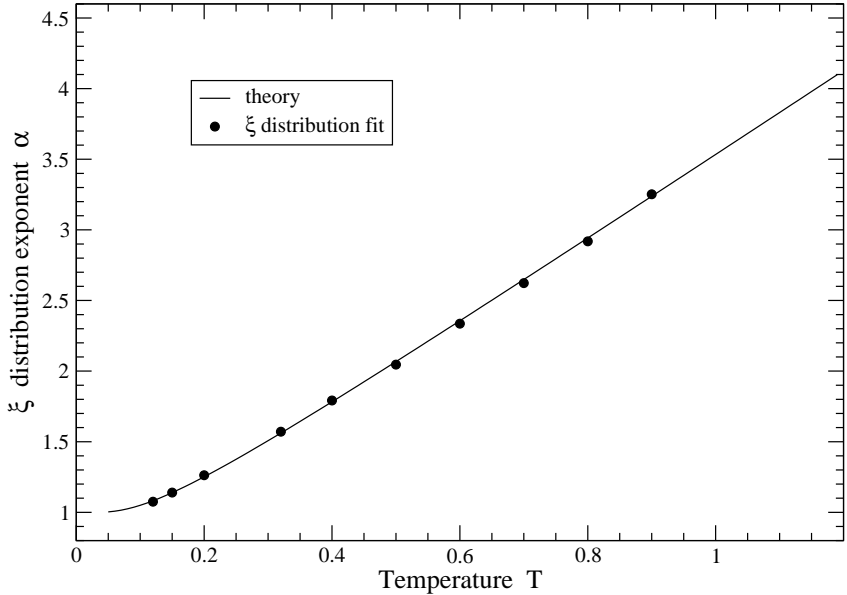}
  \caption{The critical exponent $\aa$ for the $\xi$-distribution defined in eq(\ref{expnt})~.}
  \label{FIG1}
\end{figure}

In Fig \ref{FIG1} we show results for the parameter choice $\ll=a=1$~. Using a Sturm sequencing technique 
we determine the values of $p$ for which $H_p$ acquires a zero eigenvalue for a range of temperature
$T=1/\bb$~. For sufficiently large $T$ the critical value $p_c$ is greater than 2, thus permitting the variance of 
$\xi$ to exist. However as the temperature is lowered there is then a critical value $T_c$
below which $p_c$ is less than 2, and and for those values of $T$ the variance of $\xi$ diverges. 
In the present model $T_c=0.473$~.

\subsubsection{\label{dmcritp2} Simulation for Double Minimum}

The stochastic differential equations for our choice of parameters become
\be
\fr{dx}{dt}=-x(x^2-1)+W(t)~~,
\label{dminsde1}
\ee
where $\la W(t)W(t')\ra=2T\dd(t-t')$~, and
\be
\fr{d\xi}{dt}=-(3x^2-1)\xi+1~~.
\label{dminsde2}
\ee
Using the numerical integration techniques outlined in section \ref{simul}, we can obtain 
from a sample of $N\simeq 10^6$ particles a measurement of $D(\xi,p)$ as a function
of $\xi$~. The case $p=1$ allows us to extract $\aa$ and $\chi=\bar{\xi}$, the susceptibility, by fitting the 
asymptotic form in eq(\ref{Dasymp}) to the results of the simulation over an appropriate range
of $\xi$~. 

\begin{figure}[t]
   \centering
  \includegraphics[width=0.65\linewidth]{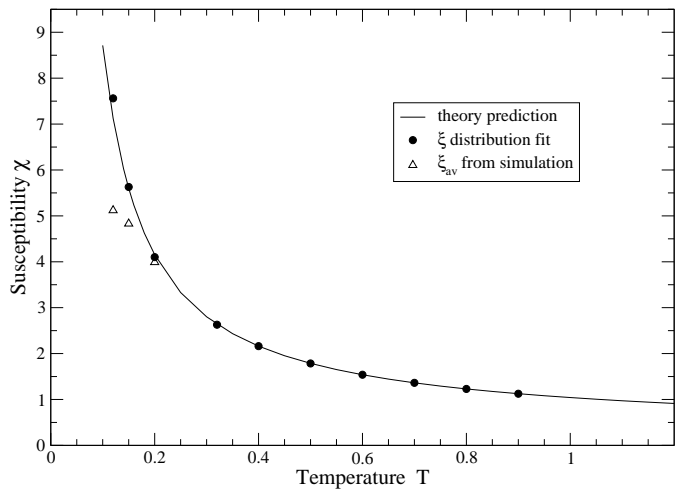}
  \caption{The susceptibility $\chi$ for the double minimum profile defined in eq(\ref{dminprof})~.}
  \label{FIG2}
\end{figure}

The variance of $x$ can be evaluated by means of a direct numerical evaluation of the
integral
\be
\la x^2\ra=\fr{1}{Z}\int dx x^2e^{-\bb\phi(x)}~~.
\label{numvar}
\ee
Then using the Einstein relation, $\chi=\xib=\bb\la{x^2}\ra$, we obtain an
evaluation of the the susceptibility.  In Fig \ref{FIG2} we show that our simulation results
compare well with this evaluation, even down to temperatures well below $T_c$~. 

Of course we could extract $\bar{\xi}$ directly from the simulation data. The results coincide 
with those obtained by the more elaborate fitting procedure outlined above for sufficiently
high values of the temperature. However for $T<0.2$ a straight average estimate yields
results, indicated in Fig \ref{FIG2} by open triangles, that are too low. The elucidation
of this inconsistency lies in the manner in which the power law tail of the $\xi$-distribution
is achieved, namely by the appearance of a sequence of large spikes in the evolution
of the variable $\xi$ as a function of time in the simulation \cite{DEAN}. These spikes, which
are essential for establishing the power law tail of the $\xi$-distribution, are increasingly 
prominent  and increasingly sparse as the temperature, $T$, is lowered. Achieving an accurate 
representation of the spikes from the numerical integration of the stochastic differential equations
is increasingly difficult as $T$ is reduced below $T_c$ requiring as it does, a longer and 
longer equilibriation time in order to establish the tail of the $\xi$-distribution. The result 
is that the distribution is underestimated at large $\xi$ yielding a value for the susceptibility, 
$\chi=\bar{\xi}$ that is too low. Our results show however that using the fitting procedure described 
above, it is possible to obtain an estimate of $D(\xi,1)$, over a range appropriately restricted at 
small and large $\xi$, that is sufficiently accurate to permit the extraction of a good estimate of $\chi$~. 

\begin{figure}[t]
   \centering
  \includegraphics[width=0.8\linewidth]{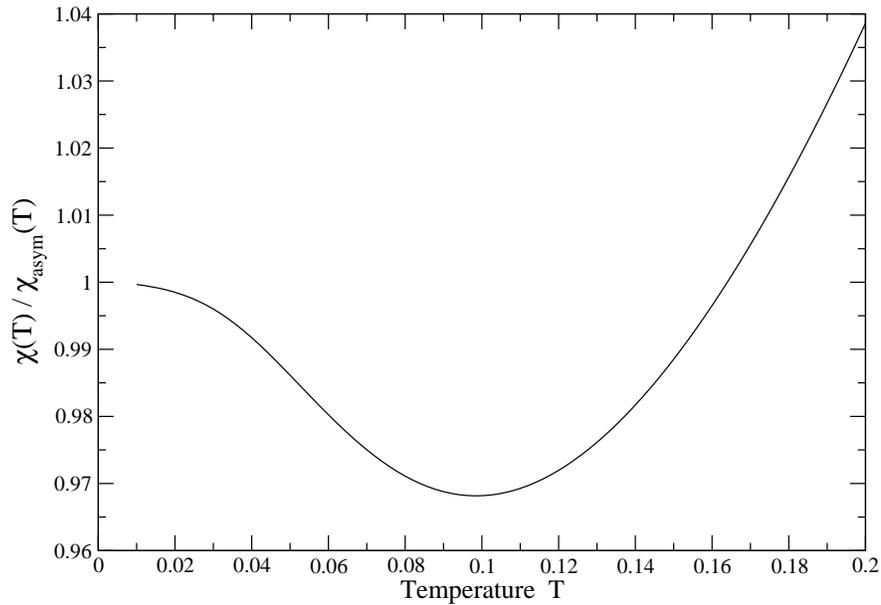}
  \caption{The Ratio of full susceptibility to asymptotic form at low $T$ for the double minimum profile
                       defined in eq(\ref{dminprof})~.}
  \label{FIG3}
\end{figure}

From eq(\ref{numvar}) we find immediately the high $\bb$ limit of the susceptibility
to be
\be
\chi=\xib=\bb(a^2-\fr{1}{\bb}\fr{1}{\ll a^2})~~.
\ee
This coincides with the specialisation of eq(\ref{variance}) to one dimension.

For our choice of parameters
\be
\chi=\fr{1}{T}-1~~.
\ee
In Fig \ref{FIG3} we show the ratio of the numerical evaluation of $\chi$ to the above asymptotic form
as a function of temperature for $T<0.2$. It correctly approaches unity at $T=0$ with zero slope.
The correction to the asymptotic form is small for $T<0.02$~. As indicated above it is difficult
to perform the stochastic simulation at these low $T$ values.

\subsubsection{\label{dmdecay1} Decay Rate for Double Minimum}

We can calculate the low $T$ estimate for the decay rate of an asymmetrical initial
$x$-distribution using the electromagnetic analogy explained above. This
requires a calculation of the resistance $R$ between the minima. 
From eq(\ref{resist1d}) we find for our model
\be
R=\sqrt{\fr{2\pi\bb}{\ll a^2}}\fr{1}{P_0(0)}
             =\sqrt{\fr{2\pi\bb}{\ll a^2}}Z\exp\left\{\fr{1}{4}\bb\ll a^4\right\}~~,
\label{res1_d}
\ee
where the partition function $Z$ is given by
\be
Z=\int^\infty_{-\infty}dxe^{-\bb\phi(x)}~~.
\ee
When $\bb$ is very large we can compute $Z$ from the dominant contributions 
coming from the neighbourhoods of the two minima. Both contributions are equal
so
\be
Z=2\int dx e^{-\bb\ll a^2(x-a)^2}=2\sqrt{\fr{\pi}{\bb\ll a^2}}~~.
\ee
We have also
\be
P_0(x)=w_1^2P_0^{(1)}(x)+w_2^2P_0^{(2)}(x)~~,
\ee
where $w_1^2=w_2^2=1/2$ and
\be
P_0^{(1,2)}(x)=\sqrt{\fr{\bb\ll a^2}{\pi}}e^{-\bb\ll a^2(x\mp a)^2}~~.
\ee
We note that $\phi''(0)=-\ll a^2$ and that
\be
P_0(0)=\fr{1}{Z}e^{-\bb\ll a^4}~~.
\ee
Applying the electrical analogy we see that the voltages at the two minima 
satisfy the equations
\be
\fr{1}{2}\fr{\d V_1}{\d t}=-\fr{1}{R}(V_1-V_2)~~,
\ee
and 
\be
\fr{1}{2}\fr{\d V_2}{\d t}=-\fr{1}{R}(V_2-V_1)~~.
\ee
where R is given by eq(\ref{res1_d})~.
\be
R=\sqrt{\fr{2\pi\bb}{\ll a^2}}\fr{1}{P_0(0)}~~.
\ee

It follows that the rate of decay of the voltage difference is proportional to $e^{-\nu t}$
where 
\be
\nu=\fr{4}{R}=\fr{\sqrt{2}\ll a^2}{\pi}e^{-\fr{1}{4}\bb\ll a^4}
\label{decind}
\ee
The eigenvalue of the supersymmetric Hamiltonian is of course $\bb\nu$~.

\begin{figure}[t]
   \centering
  \includegraphics[width=0.8\linewidth]{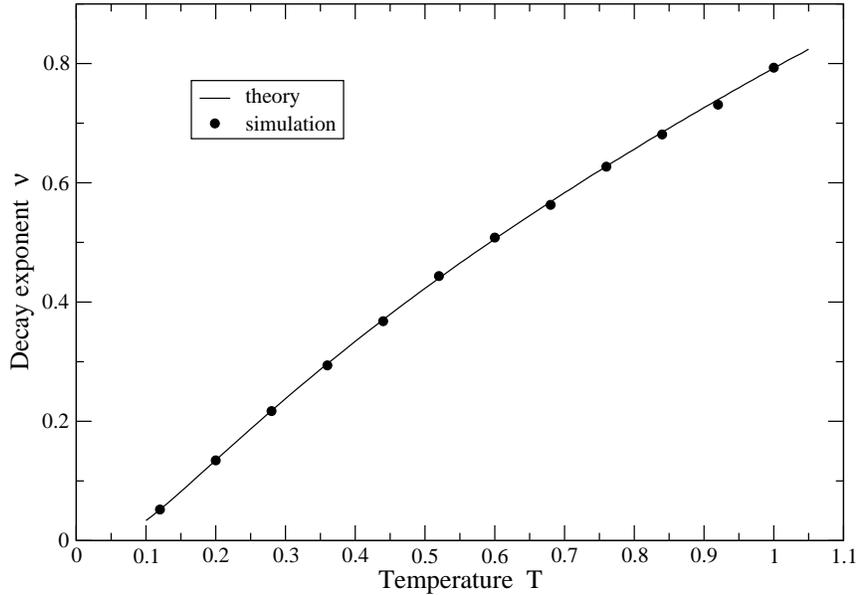}
  \caption{A comparison of the decay exponent $\nu$ for the double minimum profile defined in eq(\ref{dminprof})
    calculated theoretically from the Einstein relation and from a simulation numerically integrating
the stochastic differential equations, eq(\ref{dminsde1})~.}
  \label{FIG4}
\end{figure}

In Fig \ref{FIG4} we show the results of the simulation (for the case $\ll=a=1$). They compare well
with the direct evaluation obtained from the Sturm sequencing evaluation of the 
lowest non-zero eigenvalue of $H_0$~. For this choice of parameters the asymptotic result
becomes
\be
\nu=\fr{\sqrt{2}}{\pi}e^{-1/(4T)}~~.
\ee
For the temperature range $T\le 0.2$ this formula yields results very close to the 
simulation and numerical evaluation shown in Fig \ref{FIG4}.

\subsection{Three Minimum Profile}

As a matter of interest we work through the theoretical calculations for a more complicated 
example for which $\phi(x)$ has three minima and two maxima. We choose
\be
\phi(x)=\fr{\ll}{4}x^2(x^2-a^2)^2~~.
\ee
The minima are at the points $x=0,\pm a$, and the maxima are at $x=\pm a/\sqrt{3}$.
We have the results $\phi(\pm a/\sqrt{3})=\ll a^4/27$ and $\phi''(0)=\ll a^4/2$,
$\phi''(\pm a)=2\ll a^4$ together with $\phi''(\pm a/\sqrt{3})=-2\ll a^4/3$.

If we label the minima at $x=\pm a$ by 1 and 2, and the minimum at $x=0$ by 0 then we
easily find in the high $\bb$ limit
\be
Z_1=Z_2=\sqrt{\fr{\pi}{\bb\ll a^4}}~~,
\ee
and
\be
Z_0=\sqrt{\fr{4\pi}{\bb\ll a^4}}~~.
\ee
Hence 
\be
Z=Z_1+Z_2+Z_0=4\sqrt{\fr{\pi}{\bb\ll a^4}}~~,
\ee
and therefore $w_1^2=w_2^2=1/4$ and $w_0^2=1/2$~.

The resistance between the origin and each of the other two minima is $R$ where
\be
R=\fr{4\sqrt{3}\pi}{\ll a^4}e^{\fr{\bb\ll a^4}{27}}~~.
\ee
If the voltages at the minima are $v_1$, $V_2$ and $V_0$ then the electrical analogy
tells us that
\be
\fr{1}{4}\fr{\d V_1}{\d t}=-\fr{1}{R}(V_1-V_0)~~,
\ee
\be
\fr{1}{4}\fr{\d V_2}{\d t}=-\fr{1}{R}(V_2-V_0)~~,
\ee
and
\be
\fr{1}{2}\fr{\d V_0}{\d t}=-\fr{1}{R}(2V_0-V_1-V_2)~~.
\ee
It follows that equations for the eigencombinations are
\be
\fr{\d}{\d t}(V_1+V_2-2V_0)=-\fr{8}{R}(V_1+V_2-2V_0)~~,
\ee
and
\be
\fr{\d}{\d t}(V_1-V_2)=-\fr{4}{R}(V_1-V_2)~~.
\ee
The two decay exponents are then 
\be
\nu_1=\fr{2\ll a^4}{\sqrt{3}\pi}e^{-\fr{\bb\ll a^4}{27}}~~,
\ee
and
\be
\nu_2=\fr{\ll a^4}{\sqrt{3}\pi}e^{-\fr{\bb\ll a^4}{27}}~~.
\ee

The susceptibility can be computed from eq(\ref{variance}) to be
\be
\xib=\fr{1}{2}(a^2+\fr{1}{4\bb\ll a^4})~~.
\ee

\section{\label{twodex}Two-Dimensional Example}

We can illustrate some of the results of general theory by means of two-dimensional
models. 

\subsection{\label{blind}Blind Saddle}

In \cite{KURCHAN1} T\v{a}nase-Nicola and Kurchan introduce a simple
model that they refer to as a {\it blind saddle}. It is constructed as a
two-dimensional tilted Mexican hat for which
\be
\phi(x)=\fr{1}{4}\ll(x^2+y^2-a^2)^2-Jx~~.
\label{blindsaddle}
\ee
In this case, for sufficiently small $J$, $\phi(x)$ has one minimum near $x=a$ and 
$y=0$, one saddle with Morse index one near $x=-a$ and $y=0$ together with
a maximum, Morse index two, near the origin. All lines of steepest descent leaving
the saddle arrive at the the single minimum. The uniqueness of the minimum means
that there is no opportunity to construct states with low non-zero eigenvalues for $H_0$
in the high $\bb$ limit.  The interest of the model then, is that in the limit of 
large $\bb$ there are no eigenstates in $\Hc^{(+)}_0$ with exponentially small eigenvalues. 
The lowest eigenstates in $\Hc^{(+)}_0$ are all of $O(\bb)$~. 
However in $\Hc_1^{(+)}$ there is such a low-lying eigenstate which we associate with a 
magnetic flux loop $B$ passing round the circuit created by the lines of 
steepest descent joining the saddle to the minimum.

To detect this low lying state we must first create it. Subsequently we let it die
away and measure the exponential decay of an appropriate observable, namely
$\xib_n$~. This yields the value of the  decay exponent.

To create the state we integrate the stochastic differential equation, eq(\ref{diff1})
together with the associated slave equation, eq(\ref{slave2}) in which $f_{na}(x)$
is chosen to have a magnetic character. That is 
\be
Qf_{na}(x)a^\dag_n\psi_0=0~~.
\ee
Using the fact that $Q$ commutes with $H$, we see from eq(\ref{diff8a}) that the form of $\psi_a(x)$ 
in the limit of large time, also has the property that
\be
Q\psi_a=0~~.
\ee
This guarantees that $\psi_a\in\Hc_1^{(+)}$ and therefore corresponds to
a state with a loop of flux as described in subsection \ref{onefer}~. We can solve eq(\ref{diff8a}) 
for $\psi_a(x)$ by introducing an eigenfunction basis, $\{\om_\aa\}$, for $\Hc_1^{(+)}$~.
\be
\psi_a=\sum_\aa\om_\aa\fr{(\om_\aa,f_{na}(x)a^\dag_n\psi_0)}{E_\aa}~~.
\ee
If now we continue the simulation by integrating the associated slave
equation with $f_{na}(x)$ set to zero, the subsequent time development
of $\psi_a$ is given by
\be
\psi_a=\sum_\aa e^{-TE_\aa t}\om_\aa\fr{(\om_\aa,f_{na}(x)a^\dag_n\psi_0)}{E_\aa}~~,
\ee
where th time $t$ is measured from the new starting point.
Consider the matrix element
\be
(f_{mb}(x)a^\dag_m\psi_0,\psi_a)=\int dx P_0(x)f_{mb}(x)F_{ma}(x)
                                           =\la f_{mb}(x)\xi_{ma}\ra~~.
\ee
It follows that 
\be
\la f_{mb}(x)\xi_{ma}\ra=
\sum_\aa e^{-TE_\aa t}\fr{(f_{mb}(x)a^\dag_m\psi_0,\om_\aa)(\om_\aa,f_{na}(x)a^\dag_n\psi_0)}{E_\aa}~~.
\label{circ}
\ee
The long time behaviour of this expectation value is governed by the most
slowly varying exponential. By measuring this  expectation value in the 
simulation we can obtain the  decay exponent. At this point a limitation of the method
is encountered. A careful analysis, not presented here, shows that the scalar product 
$(\om_\aa,f_{na}(x)a^\dag_n\psi_0)=C_{a\aa}\sqrt{E_\aa}$, where the coefficient $C_{a\aa}$
depends on the form of $f_{na}(x)$~. The explicit dependence on the eigenvalue $E_\aa$
removes the potentially enhancing denominator on the right in eq(\ref{circ}) and for
reasonable choices of $f_{na}(x)$ the coefficient $C_{a\aa}$ is sharply reduced at
low temperature. This results in the decoupling of the contribution from the lowest eigenstate 
as the temperature is lowered making it harder to pick out the relevant term. 
This barrier to easy simulation at low temperature 
is a theme running through our calculations and prevents a close comparison with
the low temperature asymptotic result as we see below.

In practice we need only a single flow field and we drop reference to 
the label $a$ from now on. We choose $f_x(x)=\phi_y(x)$ and $f_y(x)=-\phi_x(x)$~.
It is easily checked that $f_n(x)a^\dag_n\psi_0\in\Hc_1^{(+)}$~. 
For the purposes of the simulation we set $\ll=a=1$ in eq(\ref{blindsaddle}).
We have therefore
\begin{eqnarray}
\phi(x,y)&=&\fr{1}{4}(x^2+y^2-1)^2-Jx\\\nonumber
\phi_x(x,y)&=&x(x^2+y^2-1)-J~~,\\\nonumber
\phi_y(x,y)&=&y(x^2+y^2-1)~~,\\\nonumber
\phi_{xx}(x,y)&=&(3x^2+y^2-1)~~,\\\nonumber
\phi_{yy}(x,y)&=&(x^2+3y^2-1)~~.\\\nonumber
\phi_{xy}(x,y)&=&2 xy\\\nonumber
\end{eqnarray}

In order not to shift the stationary points very far from $x=0,\pm 1$ we set $J=0.02$~.
These stationary points which lie on the line $y=0$~. On the left we have the saddle at $x=-0.989846$ 
with $\phi_{xx}=\gg_1=1.939384$, $\phi_{yy}=-\gg_{||}=-0.020205$, the maximum at $x=-0.020008$ with $\phi_{xx}=\aa_1=-0.998799$,
$\phi_{yy}=\aa_2=-0.999599$, and the minimum at $x=1.009854$ with $\phi_{xx}=2.059415$, $\phi_{yy}=0.019805$~.
Of course $\phi_{xy}=0$ at all three stationary points. For future reference we note that
at the maximum we have $\phi(\hbox{max})=0.250200$ and at the saddle $\phi(\hbox{saddle})=0.019899$~.
Hence the difference in heights between maximum and saddle is $d$ where
\be
d=\phi(\hbox{max})-\phi(\hbox{saddle})=0.23030106~~.
\ee
Combining eq(\ref{resist2d}) and eq(\ref{floop2d}) we find the asymptotic estimate for the
decay index
\be
\nu_{\hbox{asym}}=\fr{1}{2\pi}\sqrt{\fr{\aa_1\aa_2\gg_1}{\gg_{||}}}e^{-d/T}~~.
\ee
Using the numbers appropriate to the simulation we have
\be
\nu_{\hbox{asym}}=ge^{-d/T}~~,
\label{asympnu}
\ee
where $g=1.558012$~.
A comparison between the asymptotic formula and the simulation results is 
shown in Table \ref{TAB1}. 

Also recorded is our evaluation of the decay index as $\nu_{\hbox{th}}=TE$ where $E$
is the lowest eigenvalue of $H$ in the sector $\Hc^{(-)}_2$~. Because we are in two
dimensions this is the lowest eigenvalue of the simple Schr\"odinger Hamiltonian $\tilde{H}_0$
obtained from $H_0$ by making the replacement $\phi(x)\rightarrow -\phi(x)$~. We use
the lattice field theory method of reference \cite{CDH2} to evaluate this eigenvalue. We see that this evaluation 
compares favourably with the outcome of the simulation, particularly at the higher
values of temperature. The estimate from the asymptotic formula is less good, though
not totally out of line with the other results. Our experience with the one dimensional problem
suggests that we would require values of temperature somewhat lower than 0.1 to
see agreement with the other methods of evaluation. However as we have indicated above
these methods are not easily applied for temperatures in this range. With these provisos
we feel that the theory and simulation are in encouraging agreement. It would be
interesting to attempt theoretical evaluations of $\nu$ at the higher temperatures 
based on an approximation scheme that exploits the application of the minimum dissipation 
principle to appropriately chosen field distributions that are less extreme in form
than the very narrow wires and surfaces used in the very low $T$ limit.

\begin{table}
\centering
\begin{tabular}{|c|c|c|c|}\hline
&&&\\
~~$T$~~   &$\nu_{\hbox{sim}}$&$\nu_{\hbox{th}}$&$\nu_{\hbox{asym}}$\\
&&&\\\hline
0.4&0.8813(5)&0.884(2)&0.8760\\
0.3&0.6449(5)&0.651(1)&0.7230\\
0.2&0.3900(6)&0.392(1)&0.4925\\
0.18&0.3411(5)&0.3386(7)&0.4334\\
0.16&0.283(1)&0.2835(7)&0.3693\\
0.10&0.110(4)&0.1178(5)&0.1557\\
0.09&0.084(3)&0.0946(5)&0.1206\\\hline
\end{tabular}
\caption{Comparison of values of the decay index $\nu$ obtained
by simulation $(\nu_{\hbox{sim}})$, lowest eigenvalue in $\Hc^{(-)}_2$ $(\nu_{\hbox{th}})$
and the asymptotic estimate $(\nu_{\hbox{asym}})$~.}
\label{TAB1}
\end{table}

\subsection{\label{fourmin} Four Minimum Model}
Finally we illustrate our methods with a model that has four minima.
We choose $\phi$ have the form
\be
\phi=\phi(x,y)=\fr{1}{4}\ll(x^2-a^2)^2+\fr{1}{4}\mu(y^2-a^2)^2+\fr{1}{2}\ka x^2y^2~~,
\ee
where $x$ and $y$ are the coordinates of the diffusing particle. The relevant derivatives
are
\begin{eqnarray}
\phi_x&=&\ll x(x^2-a^2)+\ka xy^2~~,\\\nonumber
\phi_y&=&\mu y(y^2-a^2)+\ka x^2y~~,\\\nonumber
\phi_{xx}&=&\ll(3x^2-a^2)+\ka y^2~~,\\\nonumber
\phi_{yy}&=&\mu(3y^2-a^2)+\ka x^2~~.\\\nonumber
\phi_{xy}&=&2\ka xy\\\nonumber
\end{eqnarray}
The stationary points satisfy
\begin{eqnarray}
x[\ll(x^2-a^2)+\ka y^2]&=&0~~,\\\nonumber
y[\mu(y^2-a^2)+\ka x^2]&=&0~~.\\\nonumber
\end{eqnarray}
We have then the following stationary points. 
\begin{itemize}
\item[(1)] A maximum at $x=y=0$ with $\phi_{xx}=-\ll a^2$, $\phi_{yy}=-\mu a^2$,~~$\phi_{xy}=0$~.
\item[(2)] Saddles with Morse index 1 at $x=0$,~~$y=\pm a$ with
$\phi_{xx}=-(\ll-\ka)a^2$,~~$\phi_{yy}=2\mu a^2$,~~$\phi_{xy}=0$~~and at
$y=0$,~~$x=\pm a$ with $\phi_{xx}=2\ll a^2$,~~$\phi_{yy}=-(\mu-\ka)a^2$,~~$\phi_{xy}=0$~.
\item[(3)] Four minima at the points $x=\pm\sqrt{\mu(\ll-\ka)/(\ll\mu-\ka^2)}a$ and 
$y=\pm\sqrt{\ll(\mu-\ka)/(\ll\mu-\ka^2)}a$ with $\phi_{xx}=2a^2\ll\mu(\ll-\ka)/(\ll\mu-\ka^2)$
and $\phi_{yy}=2a^2\ll\mu(\mu-\ka)/(\ll\mu-\ka^2)$ and
$\phi_{xy}=\pm 2a^2\ka\sqrt{\ll\mu(\ll-\ka)(\mu-\ka)}/(\ll\mu-\ka^2)$~.
\end{itemize}
For case (3) the discriminant of the Hessian at each minimum is $\DD$ where
\be
\DD=4a^4\fr{\ll\mu(\ll-\ka)(\mu-\ka)}{\ll\mu-\ka^2}~~.
\ee
We see that these stationary points are indeed minima if $0\le\ka<\hbox{min}(\ll,\mu)$
since this guarantees that $\DD$, $\phi_{xx}$ and $\phi_{yy}$ are all positive. The eigenvalues of the Hessian matrix are 
$$
\fr{a^2}{\ll\mu-\ka^2}\left\{\ll\mu(\ll+\mu-2\ka)\pm\sqrt{\ll^2\mu^2(\ll-\mu)^2+4\ka^2\ll\mu(\ll-\ka)(\mu-\ka)}\right\}
$$

\subsubsection{\label{decay}Decay Indices}

We first calculate the decay indices in the high $\bb$-limit.
In this limit the partition function is
\be
Z=\fr{8\pi}{\bb}\fr{1}{\sqrt{\DD}}\exp\left\{-\fr{\bb a^4}{4}\fr{\ka(2\ll\mu-\ka(\ll+\mu))}{\ll\mu-\ka^2}\right\}~~.
\ee


To calculate the low lying eigenvalues in this limit we first evaluate the
resistances between adjacent minima. In the present model there are two
resistances namely, $R_x$, appropriate to paths of steepest descent passing 
in the $x$-direction through the saddles at $x=0$, $y=\pm a$, and $R_y$,
appropriate to paths passing in the $y$-direction through the saddles at $y=0$, $x=\pm a$~.
Using the result in eq(\ref{resist2d}) we find
\be
R_x=\fr{8\pi}{\sqrt{\DD}}\sqrt{\fr{2\mu}{\ll-\ka}}\exp\left\{\fr{\bb a^4}{4}\fr{\mu(\ll-\ka)^2}{\ll\mu-\ka^2}\right\}~~,
\ee
and
\be
R_y=\fr{8\pi}{\sqrt{\DD}}\sqrt{\fr{2\ll}{\mu-\ka}}\exp\left\{\fr{\bb a^4}{4}\fr{\ll(\mu-\ka)^2}{\ll\mu-\ka^2}\right\}~~.
\ee
For convenience we label the minima with $r=1,2,3,4$ in a clockwise direction starting in the positive quadrant.
The symmetry of the model is sufficient to ensure that the capacitances at all four 
minima are equal. That is $w_r^2=1/4$~.
From eq(\ref{volts}) we obtain the equations of
current flow as
\begin{eqnarray}
\fr{1}{4}\fr{\d V_1}{\d t}&=-(V_1-V_2)/R_y-(V_1-V_4)/R_x~~,\\\nonumber
\fr{1}{4}\fr{\d V_2}{\d t}&=-(V_2-V_3)/R_y-(V_2-V_1)/R_x~~,\\\nonumber
\fr{1}{4}\fr{\d V_3}{\d t}&=-(V_3-V_4)/R_y-(V_3-V_2)/R_x~~,\\\nonumber
\fr{1}{4}\fr{\d V_4}{\d t}&=-(V_4-V_3)/R_y-(V_4-V_1))/R_x~~.\\\nonumber
\end{eqnarray}
The three eigenmodes of the form $V_r=U_re^{-pt}$ with $p\ne 0$ are
\be
\begin{array}{ccc}
          U_1=U_2=U&U_3=U_4=-U&p=8/R_x~~,\\
          U_1=U_4=U&U_2=U_3=-U&p=8/R_y~~,\\
          U_1=U_3=U&U_2=U_4=-U&p=(8/R_x+8/R_y)~.
\end{array}
\ee

We conclude that the three  decay exponents associated with the 
subspaces $\Hc_0^{(+)}$ and $\Hc_1^{(-)}$ are $\nu_1=8/R_x$, $\nu_2=8/R_y$ and $\nu_3=(8/R_x+8/R_y)$~.
Note that these results at low $T$ imply that $\nu_3=\nu_1+\nu_2$~. However we do not expect this simple
relation to hold for general values of the temperature.

To evaluate these decay exponents for values of $T$ above the low
temperature limit we performed a simulation by numerically integrating the 
stochastic equations of motion. Our procedure was to equilibriate the 
system and then to fold over the resulting configuration by reflecting the position
of each particle in appropriate axes so that it ended up in the positive $(x,y)$-quadrant,
the remaining three quadrants then being empty. We then restarted the evolution of the 
system and measured the expectation values $\la x\ra$, $\la y\ra$ and $\la xy\ra$
as a function of time. The long time behaviour of each of these three quantities yields respectively
the decay exponents $\nu_1$, $\nu_2$ and $\nu_3$~.  In Table \ref{TAB2} we show the results 
in the case $a=1.0$, $\ll=1.0$ and $\mu=0.5$ for a number of values of $\ka$ and $T$~.
The case $\ka=0.0$ corresponds to a situation in which motion in the $x$ and $y$ directions
are independent. This serves as a check on our procedures since under these circumstances
we expect that $\nu_3=\nu_1+\nu_2$ at all temperatures.
Within errors Table\ref{TAB2} shows that this is indeed the case. For $\ka\ne 0.0$ we 
expect this outcome only at low temperatures, as indicated above. At the temperatures listed in Table \ref{TAB2}
the result does not hold as indeed we anticipated.

\begin{table}
\centering
\begin{tabular}{|c|c|c|c|c|}\hline
&&&&\\
$\ka$&~~$T$~~   &$\nu_1$&$\nu_2$&$\nu_3$\\
&&&&\\\hline
0.0&0.8&0.6573(5)&0.5704(6)&1.222(5)\\
&0.6&0.5065(4)&0.4626(6)&0.9632(2)\\\hline
0.2&0.8&0.7598(6)&0.6612(4)&1.566(3)\\
&0.6&0.5847(3)&0.5452(3)&1.239(2)\\\hline
0.3&0.8&0.8000(7)&0.6985(4)&1.732(4)\\
&0.6&0.6174(3)&0.5792(4)&1.373(4)\\\hline
\end{tabular}
\caption{Simulation results for the decay indices $\nu_1$, $\nu_2$ and $\nu_3$
for the Four Minimum model.}
\label{TAB2}
\end{table}

General theory tells us that the dimension of the low lying subspace
in $\Hc_1$ is equal to the number of saddles. As we have seen three of the
basis states are in $\Hc_1^{(-)}$ so a remaining low lying state lies in
$\Hc_1^{(+)}$ with a corresponding supersymmetric partner in $\Hc_2^{(-)}$~.
In our electromagnetic analogy this state is associated with a loop of magnetic flux
(in the present example confined to the two-dimensional surface)
that runs through all the minima
and surrounds the maximum of $\phi$ at the point $(0,0)$~. 

The loop resistance
is given by
\be
R_L=2(R_x +R_y)
\ee
From eq(\ref{floop2d}) we can conclude that the associated  decay exponent is 
\be
\nu_4=\fr{\sqrt{2}a^2}{\pi}\left\{\mu\sqrt{\fr{\ll}{\ll-\ka}}\exp\left\{-\fr{\mu\bb a^4}{4}\right\}
                         +\ll\sqrt{\fr{\mu}{\mu-\ka}}\exp\left\{-\fr{\ll\bb a^4}{4}\right\}\right\}~~.
\ee

The decay index $\nu_4$ is associated with the subspaces $\Hc_1^{(+)}$ and $\Hc_2^{(-)}$~.
We evaluate it at higher temperatures by using the same procedure {\it mutatis mutandis} as for the 
blind saddle. The results are shown in Table\ref{TAB3}~. Also shown for comparison is a theoretical
evaluation of the index, $\nu_{4(\hbox{th})}$, obtained using the Schr\"odinger problem
with $\phi(x)\rightarrow-\phi(x)$ as was done for the blind saddle. The two sets of results
compare reasonably well providing good evidence that the numerical methods work reliably. 
It is worth noting that for the case $\ka=0.0$ we $\nu_3=\nu_4$, as expected.

For practical reasons both the simulation and the theoretical Schr\"odinger calculation
of $\nu_4$ cannot be carried out at a temperature sufficiently low that a comparison
with the the asymptotic calculation is possible. 
\begin{table}
\centering
\begin{tabular}{|c|c|c|c|}\hline
&&&\\
$\ka$&~~$T$~~&$\nu_4$&$\nu_{4(\hbox{th})}$\\
&&&\\\hline
0.0&0.8&1.354(1)&1.240(2)\\
&0.6&0.954(3)&0.965(2)\\\hline
0.2&0.8&1.313(2)&1.317(2))\\
&0.6&1.031(1)&1.039(2))\\\hline
0.3&0.8&1.354(1)&1.347(2)\\
&0.6&1.068(1)&1.065(1)\\\hline
\end{tabular}
\caption{Simulation results for the decay index $\nu_4$
for the Four Minimum model together with a theoretical evaluation $\nu_{4(\hbox{th})}$.}
\label{TAB3}
\end{table}

\subsubsection{\label{suscep4m}Susceptibilities}

We first calculate the susceptibility at low temperature.
From eq(\ref{variance}) we see that there are two types of contribution
from each of the minima. The overall mean position of the particle 
is the origin in this model and the capacity of each minimum is $1/4$.
We obtain from minimum $1$ contributions to $\SS_{xx}$, $\SS_{yy}$ and $\SS_{xy}$ of the form
$\fr{1}{4}(\xb^{(1)})^2$, $\fr{1}{4}(\yb^{(1)})^2$ and $\fr{1}{4}(\xb^{(1)})\yb^{(1)}$ where $\xb^{(1)}$ 
and $\yb^{(1)}$ refer to the position of the minimum modified 
by terms that are $O(1/\bb)$~. This modification is computed by setting
\begin{eqnarray}
x&=&a\sqrt{\fr{\mu(\ll-\ka)}{\ll\mu-\ka^2}}+x'~~,\\
y&=&a\sqrt{\fr{\ll(\mu-\ka)}{\ll\mu-\ka^2}}+y'~~.
\end{eqnarray}
We then consider $x'$ and $y'$ to be small quantities and evaluate $\phi(x,y)$
in the neighbourhood of the minimum as
\be
\phi(x,y)=\phi(\hbox{minimum})+\hbox{quadratic}~~.
\ee
where
\begin{eqnarray}
\hbox{quadratic}&=&\fr{1}{2}(\phi_{xx}x'^2+\phi_{yy}y'^2+2\phi_{xy}x'y')+\hbox{cubic}~~,\\
\hbox{cubic}&=&\fr{1}{6}(\phi_{xxx}x'^3+\phi_{yyy}y'^3+3\phi_{xxy}x'^2y'+3x'y'^2)~~.
\end{eqnarray}
If we retain this cubic term in $\phi$ in evaluating the local average of $x$ and $y$ at minimum $1$ 
we find to $O(1/\bb)$
\begin{eqnarray}
\xb^{(1)}&=&a\sqrt{\fr{\mu(\ll-\ka)}{\ll\mu-\ka^2}}
       \left(1-\fr{1}{4\bb a^4}\fr{3\ll^2\mu(\mu-\ka)+2\ka^3(\ll-\ka)}{\ll\mu(\ll-\ka)^2(\mu-\ka)}\right)~~,\\ 
\yb^{(1)}&=&a\sqrt{\fr{\ll(\mu-\ka)}{\ll\mu-\ka^2}}
       \left(1-\fr{1}{4\bb a^4}\fr{3\ll\mu^2(\ll-\ka)+2\ka^3(\mu-\ka)}{\ll\mu(\ll-\ka)(\mu-\ka)^2}\right)~~.
\end{eqnarray}
The average positions at the other minima are obtained by applying the same 
modifying factors.

\begin{figure}[t]
   \centering
  \includegraphics[width=0.8\linewidth]{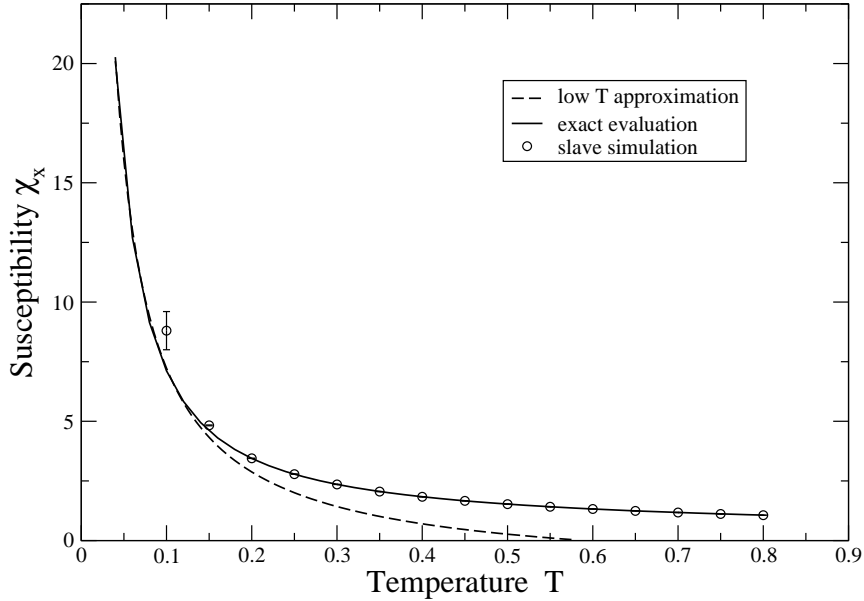}
  \caption{The susceptibility $\chi_x=\SS_{xx}$ for the Four Minimum Model calculated in the low temperature
limit, exactly from the the Einstein relation and from the stochastic and slave equations.}
  \label{FIG5}
\end{figure}

The second set of contributions from eq(\ref{variance}) are obtained by forming the local averages at
minimum $1$ of $x'^2$, $y'^2$ and $x'y'$ keeping only the quadratic approximation for $\phi$~.
We find
\begin{eqnarray}
\overline{x'^2}&=&\fr{1}{2\bb a^2}\fr{1}{\ll-\ka}~~,\\
\overline{y'^2}&=&\fr{1}{2\bb a^2}\fr{1}{\mu-\ka}~~,\\
\overline{x'y'}&=&-\fr{1}{2\bb a^2}\fr{\ka}{\sqrt{\ll\mu(\ll-\ka)(\mu-\ka)}}~~.
\end{eqnarray}
The the other minima yield identical results except for a sign change in the $\overline{x'y'}$
from minima $2$ and $3$~. The outcome from adding all the contributions from each of the minima is that 
$\SS_{xy}$ vanishes and
\begin{eqnarray}
\SS_{xx}&=&a^2\fr{\mu(\ll-\ka)}{\ll\mu-\ka^2}\left(1
  -\fr{1}{2\bb a^2}\fr{2\ll^2\mu(\mu-\ka)+\ka^2(\ll\mu-\ka^2+\ka(\ll-\ka))}{\ll\mu(\ll-\ka)^2(\mu-\ka)}\right)~~.
\label{suslowT1}\\
\SS_{yy}&=&a^2\fr{\ll(\mu-\ka)}{\ll\mu-\ka^2}\left(1
  -\fr{1}{2\bb a^2}\fr{2\ll\mu^2(\ll-\ka)+\ka^2(\ll\mu-\ka^2+\ka(\mu-\ka))}{\ll\mu(\ll-\ka)(\mu-\ka)^2}\right)~~.
\label{suslowT2}
\end{eqnarray}
We show in Fig \ref{FIG5} this low-$T$ asymptotic dependence of the susceptibilities for $a=1.0$,
$\ll=1.0$ and $\mu=0.5$~.
For the same parameter values
we can obtain the susceptibilities from a direct simulation of the stochastic 
differential equations. Guided by our experience with the one-dimensional problem
we extract the results by fitting an appropriate power law form to the 
measured cumulative distributions for $\xi_{xx}$ and $\xi_{yy}$~. The results
for the susceptibility in the $x$-direction
are exhibited in Fig \ref{FIG5} in comparison with those predicted
using the Einstein relation from an evaluation by direct numerical integration
of $\la x^2\ra$ and the asymptotic low $T$ results from eq(\ref{suslowT1}).

There is good agreement between the simulation and the direct numerical valuation
for temperatures $T>0.15$~. For lower values of $T$ here is difficulty in achieving
good simulation results for the same reason as in the one dimensional case, namely that
the distributions for $\xi_{xx}$ and $\xi_{yy}$ acquire such slowly descending power law tails 
that the do not have finite variances. However the asymptotic estimate does agree well with
the direct numerical evaluation at low values of $T$~.

\begin{table}
\centering
\begin{tabular}{|c|c|c|c|c|c|}\hline
&&&&&\\
$\ka$&~~$T$~~&$\aa_x$&$\aa_y$&$\aa_{x\wedge y}$&$\chi_{\hbox{area}}$\\
&&&&&\\\hline
0.0&0.8&2.76(1)&4.7(3)&2.88(2)&1.835750(3)\\
&0.6&2.31(1)&3.8(1)&2.33(2)&2.8086(1)\\\hline
0.2&0.8&3.62(3)&6.2(2)&3.29(8)&1.5734(2)\\
&0.6&2.99(2)&4.79(2)&2.60(1)&2.38711(6)\\\hline
0.3&0.8&4.07(7)&6.9(6)&3.5(1)&1.47233(2)\\
&0.6&3.29(4)&5.64(9)&2.72(2)&2.22607(4)\\\hline
\end{tabular}
\caption{Simulation results for the power law tail index for the
$x$-susceptibility ($\aa_x$), $y$-susceptibility ($\aa_y$),the area 
susceptibility ($a_{x\wedge y}$) and the value of the area susceptibility }
\label{TAB4}
\end{table}

In Table \ref{TAB4} we exhibit the results from the simulation for the indices
of the power law tails at a number of values for $\ka$  and the temperature $T$~.  
The first point of note is that for the case $\ka=0$ and at both temperatures, 
the area index $\aa_{x\wedge y}$ is reasonably coincident with the smaller of 
the two linear indices namely $\aa_x$~. The explanation for this lies in the fact
that when $\ka=0$ the variables $\xi_{xx}$ and $\xi_{yy}$ evolve independently 
and the area element is just $z=\xi_{xx}\xi_{yy}$~. In the appendix we show that
a product variable such, as $z$, has in general, a power law tail with an
index that coincides with the smaller of the two indices involved in the product.
Of course when $\ka$ is no longer zero we cannot infer any such simple connection
and the results in Table \ref{TAB4} bear this out. 

In Table \ref{TAB4} we have included 
the area susceptibility for the relevant values of $\ka$ and $T$~. We found that obtaining
a good outcome for the simulation at much lower temperatures rather difficult given
the size, already considerable, of the statistical sample. Although we can in principle
compute the low $T$ limit of the area susceptibility by generalising previous calculations
for the linear susceptibility we do not pursue this point here. Since it is not possible
to construct an equivalent to the direct numerical evaluation the linear susceptibility through
the application of the Einstein relation we cannot carry out a numerical comparison. 
Nevertheless an exploration of such higher susceptibilities remains an interesting topic of further study.

\section{\label{conc}Conclusions}

We have studied a stochastic system comprising a particle with degrees of freedom $\{X_n\}$
together with associated slave variables $\{\Xi_{ma}\}$ that represent infinitesimal line elements carried along
by the diffusing system. We refer to the equations satisfied by these slave variables as
slave equations. The states of the system are described by a joint probability distribution $P(x,\xi)$,
which may depend on time and which satisfies an appropriately generalised diffusion equation.
Here $\{x_n\}$ and $\{\xi_{ma}\}$ are values attained by the stochastic variables $\{X_n\}$ and $\{\Xi_{ma}\}$~. 
The slave variables can be used to construct a hierarchy of infinitesimal
areas, volumes {\it etc}, that evolve either from an initial set of (infinitesimal) displacements
of the initial conditions of the stochastic system or through alterations of trajectory
induced by external forces. In the former situation we are concerned with decay exponents that
describe the return of the system to equilibrium. In the latter we are interested in the (infinitesimal)
response of the equilibrium distribution for the system to the presence of external forces. This
response can be described by a hierarchy of susceptibilities, linear, area {\it etc}. There are 
exponents associated with these susceptibilities as a result of the fact that the probability
distributions of the associated variables have power-law tails. These exponents play a 
significant role in extracting the values of the susceptibilities from the simulation data.

The state of the system can alternatively be described in terms of appropriately 
antisymmetrized moments of the slave variables $\{\xi_{ma}\}$~. In turn these moments may be regarded as
the wavefunctions in a hierarchy of states that are the basis of a supersymmetric description
of the original system. The behaviour of the system is controlled by a supersymmetric Hamiltonian
as explained by \cite{KURCHAN1,KURCHAN2}~.  The decay exponents characterising the return
of the system to equilibrium are, up to a factor of the temperature $T$, the eigenvalues of the 
supersymmetric Hamiltonian \cite{KURCHAN1,KURCHAN2,GRAHAM}. The susceptibilities are expectation values
of antisymmetrized moments of the slave variables and are therefore related in a natural way
to the hierarchy of supersymmetric states. This supersymmetric structure is the same as that
introduced by Witten\cite{WITTEN} and investigated further by T{\v a}nase-Nicola and Kurchan \cite{KURCHAN1,KURCHAN2}~.

In order to obtain the decay exponents we follow a two pronged approach, theoretical 
on the one hand and computational on the other. 
In the theoretical approach we exploit an electromagnetic analogy and formulate a Principle of Minimum Dissipation
that provides a natural basis for computing decay exponents. It is particularly effective  
in the limit of very low temperature. It might also be made the basis, in the style of the Rayleigh-Ritz method, 
for an approximate numerical computation of exponents at higher temperatures, though we have not 
pursued this idea here. 

The computational approach we use is to some extent dependent on the number of 
degrees of freedom of the system and for simplicity we restrict our calculations to one and two degrees of
freedom although there is no barrier to working with higher dimensional systems. For one dimensional systems
we perform a direct numerical integration of the stochastic equations of motion starting with a 
non-equilibrium distribution of the variable. By measuring the  expectation value of an appropriate observable 
we can observe its ultimately exponential decay with time and measure the decay exponent directly. This works well
above a certain temperature. Unfortunately at its lower end this temperature range is just marginally above
the range of applicability of the low temperature theoretical results. However in the one dimensional case
it is simple to solve the relevant supersymmetric eigenvalue problem directly by the method of Sturm sequencing. 
This produces accurate results over a temperature range that covers both the low temperature asymptotic range and the higher
simulation range. We obtain good consistency in both cases. This confirms the efficacy of the numerical simulation
the correctness of the low temperature evaluation. 

For systems with two degrees of freedom we again carry out a simulation by integrating the stochastic differential
equations including the slave equations. Again by measuring the expectation values of appropriate variables
we can measure the exponential decay that yields the low lying eigenvalues of the supersymmetric Hamiltonian.
It should be noted however that in order to obtain the full complement of these low lying eigenvalues it is 
necessary to include slave variables that respond to behaviour in the higher fermion sector $\Hc_1^{(+)}$~. 
This is particularly true for the case of the blind saddle \cite{KURCHAN1,KURCHAN2} where the only low lying 
eigenvalue of the supersymmetric Hamiltonian is in this sector. For this model and the four minimum model 
it is not easy to compare directly with the low temperature asymptotic calculations. However qualitatively 
the results are not bad. By exploiting lattice field theory techniques \cite{CDH2} we were able to compute 
the eigenvalue appropriate to the subspace $\Hc_2^{(-)}$ which because of the supersymmetry inherent in the problem
is the same as the eigenvalue for the $\Hc_1^{(+)}$ subspace. This yielded results consistent with the numerical 
simulation.

In studying susceptibilities we again use two approaches one theoretical covering
low temperatures, the other numerical applicable at higher temperatures. The
theoretical calculation is exhibited in subsection \ref{SUSC}. It shows that 
at low temperatures the linear susceptibility has two contributions, one associated with the 
positions and occupation of the neighbourhoods of the various minima and the other associated
with a sum over the intrinsic susceptibilities of each of the minima. The numerical approach
involves integrating the stochastic equations including the slave equations. After
equilibriation we can in principle compute the mean value of $\Xi_{ma}$ directly from a sample of results 
obtained by continuing the integration procedure. At sufficiently high temperatures this works well.
However at lower temperatures, as explained in \cite{DEAN} and subsection \ref{onedex},
the probability distribution for the slave variable, $\xi$, which has a power-law tail at high 
values of $\xi$~. At sufficiently low temperatures the exponent of this
power is so small that it prevents the slave variable having a variance. This makes the
estimation of the average from the simulation impossible. However we were able to obtain
good results by compiling a running average, which does have a variance, and by fitting 
the appropriate asymptotic form to the output from the numerical simulation. We were able,
for the one dimensional case, to formulate a method for computing the exponent and
hence deduce the critical temperature below which the variance diverges. 
This amounts to constructing a
modified Hamiltonian $H_p$, associated with the moment $\bar{\xi^p}$ and then adjusting the value of $p$
until $H_p$ acquires a zero eigenvalue.  This procedure is easily carried out numerically.
In two dimensions we obtained similar results but without a method of establishing
the critical temperature theoretically. This emphasises the importance of the simulations
approach for more complicated dynamical systems. It is also possible to compute the linear susceptibility
in any number of dimensions by computing directly the variance of the position variables $X_n$
by using the Einstein relation (see section \ref{einstein}). 
We obtained good agreement between the different numerical approaches and where relevant with the theoretical
calculation. In two dimensions it is possible to compute an area susceptibility. We were able to
obtain numerical results at sufficiently high temperatures from the numerical simulation.
However we were not able to establish a useful version of the Einstein relation in this case  
in order to provide a check. Further work is also needed to investigate the theoretical 
results for the low temperature limit.

A question of interest is how the various susceptibilities might be measured.
In a simulation there is no difficulty since the various relevant quantities can be
computed in a direct manner by integrating the stochastic differential equation.
In fact a measurement of the various correlators and  decay exponents and
their inter-relationships provides a useful means of establishing the consistency
of the simulation and aids detective work on the structure of the landscape
potential function.  The experimental measurement
of the standard susceptibility is, conceptually at least, also straightforward.
One is looking at the response of the mean displacement to an external field.
An experimental measurement of higher order susceptibilities remains for the moment elusive.
However an example, in a different context, in which higher order elements, tetrads in fact,
play a significant role is in the computation of the magnetic $\aa$-effect for magnetic
field generation \cite{DRUHORG}. Finally it is worth pointing out that elucidating the structure and
evaluating the spectrum of supersymmetric quantum mechanical systems is an intrinsically
interesting and important goal.

\section*{\label{appdxA}Appendix}

We analyse here by means of a simple example the asymptotic
property of the distribution of a variable $z$ that is the product
of two independently distributed variables $u$ and $v$~. Because 
we are concerned with distributions on the range $0<u,v<\infty$
that exhibit power law behaviour for large values of the variables
we choose as the distributions $P(u,\aa)$ and $P(v,\bb)$ where
\begin{eqnarray}
P(u,\aa)&=&(\aa-1)\fr{x}{(1+x^2)^{(\aa+1)/2}}\\
P(v,\gg)&=&(\gg-1)\fr{x}{(1+x^2)^{(\gg+1)/2}}
\end{eqnarray}

The distribution for $z$ is
\be
Q(z)=\int dudv\dd(z-uv)P(u,\aa)P(v,\gg)~~.
\ee
The Mellin transform of $P(u,\aa)$ is
\be
{\tilde{P}}(l,\aa)=\int_0^\infty \fr{du}{u^{l+1}}P(u,\aa)~~,
\label{Mellin}
\ee
and the inverse relation is
\be
P(u,\aa)=\fr{1}{2\pi i}\int dl u^l{\tilde{P}}(l,\aa)~~,
\ee
and similarly for the other distributions. The $l$-integration
is along a contour in the imaginary direction with a real part
appropriately positioned. By splitting the $u$-integration range
into $0<u<1$ and $1<u<\infty$ we find explicitly
\be
{\tilde{P}}(l,\aa)=(\aa-1)\sum_{n=0}^\infty C^{-\fr{\aa+1}{2}}_n\fr{1}{2n+1-l}
                            +(\aa-1)\sum_{n=1}^\infty C^{-\fr{\aa+1}{2}}_n \fr{1}{2n+\aa+l}+\fr{\aa-1}{l+\aa}~~.
\ee
The pole at $l=-\aa$ arises from the behaviour $P(u,\aa)\simeq (\aa-1)/x^\aa$ as $x\rightarrow\infty$~.
The $l$-contour passes between this pole and that at $l=1$ which controls the behaviour at small $u$~.
Similarly ${\tilde{P}}(l,\bb)$ has a pole at $l=-\bb$~.

It is easily seen that 
\be
{\tilde{Q}}(l)={\tilde{P}}(l,\aa){\tilde{P}}(l,\gg)~~.
\ee
Therefore ${\tilde{Q}}(l)$ has poles at $l=-\aa$ and $l=-\bb$~. If $\bb>\aa$ then the 
large $z$ asymptotic behaviour will be controlled by the rightmost pole at $l=-\aa$~.
That is 
\be
Q(z)\propto \fr{1}{z^\aa}~~,
\ee
for large $z$~.
The only change to this result arises when $\bb=\aa$~. In this case $Q(l)$ has
a double pole at $l=\aa$ and the dominant behaviour at large $z$ becomes
\be
Q(z)\propto \fr{\log{z}}{z^\aa}~~.
\ee
Of course if $\bb\simeq\aa$ then it is probably more useful to include both asymptotic
contributions. However the basic result is that the distribution for $z$ has the same 
inverse power behaviour as that of the most slowly decreasing of the two distributions.

\bibliography{nsp}

\begin{thebibliography}{10}

\bibitem{WALES}
D.~J. Wales.
\newblock Energy landscapes.
\newblock {\em Cambridge University Press}, 2003.

\bibitem{PARISI}
G.~Parisi and Wu~Yongshi.
\newblock {\em Sci. Sin.}, {\bf 24}:483, 1981.

\bibitem{DDH}
I.~T. Drummond, S.~Duane, and R.~R. Horgan.
\newblock {\em Nucl. Phys. B}, {\bf 280}:25, 1987.

\bibitem{CDH}
S.~M. Catterall, I.~T. Drummond, and R.~R. Horgan.
\newblock {\em Phys. Lett.}, {\bf 254}:177, 1991.

\bibitem{DEAN}
D.~S. Dean et~al.
\newblock {\em Phys Rev E}, {\bf 70}:011101, 2004.

\bibitem{MOFFATT}
H.~K. Moffatt.
\newblock Magnetic field generation in electrically conducting fluids.
\newblock {\em Cambridge University Press}, 1979.

\bibitem{DRUHORG}
I.~T. Drummond and R.~R. Horgan.
\newblock {\em J. Fluid Mech.}, {\bf 163}:425, 1986.

\bibitem{POPE}
S.~B. Pope, P.~K. Yeung, and S.~S. Girimaji.
\newblock {\em Phys. Fluids A}, {\bf 1}:2010, 1989.

\bibitem{DRU1}
I.~T. Drummond and W.~M{\"u}nch.
\newblock {\em J. Fluid Mech.}, {\bf 215}:45, 1990.

\bibitem{DRU2}
I.~T. Drummond and W.~M{\"u}nch.
\newblock {\em J. Fluid Mech.}, {\bf 225}:529, 1991.

\bibitem{DRU3}
I.~T. Drummond.
\newblock {\em J. Fluid Mech.}, {\bf 252}:479, 1993.

\bibitem{SCH}
A.~Schkochihin, S.~Cowley, and J.~Maron.
\newblock {\em Phys. Rev. E}, {\bf 65}:016305, 2001.

\bibitem{JUNKER}
G.~Junker.
\newblock Supersymmetric methods in quantum and statistical physics.
\newblock {\em Springer Verlag, Berlin}, 1996.

\bibitem{WITTEN}
E.~Witten.
\newblock {\em J. Diff. Geom.}, {\bf 17}:661--692, 1982.

\bibitem{KURCHAN1}
S.~T{\v a}nase-Nicola and J.Kurchan.
\newblock {\em J. Stat. Phys.}, {\bf 116}:1201--1245, 2004.

\bibitem{KURCHAN2}
S.~T{\v a}nase-Nicola and J.Kurchan.
\newblock {\em Phys. Rev. Lett.}, {\bf 91}:188301--1, 2003.

\bibitem{GRAHAM}
R.~Graham.
\newblock {\em Europhysics Letters}, {\bf 5}:101--106, 1988.

\bibitem{KURCHAN3}
S.~T{\v a}nase-Nicola and J.Kurchan.
\newblock {\em J. Phys. A: Math. Gen.}, {\bf 36}:10299--10324, 2003.

\bibitem{NICOLAI1}
H.~Nicolai.
\newblock {\em Phys. Lett.}, {\bf 89B}:341--346, 1980.

\bibitem{NICOLAI2}
H.~Nicolai.
\newblock {\em Nucl. Phys.}, {\bf B176}:419--428, 1980.

\bibitem{LON}
F.~H. London.
\newblock {\em Proc Roy Soc.}, A{\bf 149}:71, 1935.

\bibitem{DEGN}
P.~G. de~Gennes.
\newblock Superconductivity of metals and alloys.
\newblock {\em Addison-Wesley}, 1989.

\bibitem{CDH2}
S.~M. Catterall, I.~T. Drummond, and R.~R. Horgan.
\newblock {\em J. Phys. A: Math. Gen.}, {\bf 24}:481, 1991.

\end{thebibliography}
\bibliographystyle{unsrt}

\end{document}